\documentclass[11pt,preprint]{aastex}
\pdfoutput=1
\usepackage{graphicx}

\let\oldfootsep=\footnotesep
\newcommand\ltsima{$\; \buildrel <\over\sim \;$}
\newcommand\simlt{\lower.5ex\hbox{\ltsima}}
\newcommand\gtsima{$\; \buildrel >\over\sim \;$}
\newcommand\simgt{\lower.5ex\hbox{\gtsima}}

\newcommand\msun {M_\odot}

\newcommand\mearth {{M_\oplus}}

\newcommand\pac{Paczy{\'n}ski }

\newcommand{\mathbold}[1]{\mbox{\boldmath $\bf#1$}}

\newcommand\muSbold{{\mathbold \mu_s}}
\shorttitle{}
\shortauthors{Bennett et al}

\begin{document}


\title{MOA-2011-BLG-262Lb: 
A Sub-Earth-Mass Moon Orbiting a Gas Giant Primary or
a High Velocity Planetary System in the Galactic Bulge }


\author{D.P.~Bennett\altaffilmark{1,M,P},
V.~Batista\altaffilmark{2,3,P,U},
I.A.~Bond\altaffilmark{4,M},
C.S.~Bennett\altaffilmark{5,6,P}, \\
D.~Suzuki\altaffilmark{7,M},
J.-P.~Beaulieu\altaffilmark{3,P},
A.~Udalski\altaffilmark{8,O},
J.~Donatowicz\altaffilmark{9},
\\
and \\
F.~Abe\altaffilmark{10},
C.S.~Botzler\altaffilmark{11},
M.~Freeman\altaffilmark{11},
D.~Fukunaga\altaffilmark{10},
A.~Fukui\altaffilmark{12},
Y.~Itow\altaffilmark{10}, 
N.~Koshimoto\altaffilmark{7},
C.H.~Ling\altaffilmark{4},
K.~Masuda\altaffilmark{10},
Y.~Matsubara\altaffilmark{10},
Y.~Muraki\altaffilmark{10},
S.~Namba\altaffilmark{7},
K.~Ohnishi\altaffilmark{13},
N.J.~Rattenbury\altaffilmark{11},
To.~Saito\altaffilmark{15},
D.J.~Sullivan\altaffilmark{14},
T.~Sumi\altaffilmark{7},
W.L.~Sweatman\altaffilmark{4},
P.J.~Tristram\altaffilmark{16},
N.~Tsurumi\altaffilmark{10},
K.~Wada\altaffilmark{7},
P.C.M.~Yock\altaffilmark{11},
\\ (The MOA Collaboration) \\
M.D.~Albrow\altaffilmark{17},
E.~Bachelet\altaffilmark{18},
S.~Brillant\altaffilmark{19},
J.A.R.~Caldwell\altaffilmark{20},
A.~Cassan\altaffilmark{3},
A.A.~Cole\altaffilmark{21},
E.~Corrales\altaffilmark{3},
C.~Coutures\altaffilmark{3},
S.~Dieters\altaffilmark{21},
D.~Dominis Prester\altaffilmark{22},
P.~Fouqu\'e\altaffilmark{18},
J.~Greenhill\altaffilmark{21},
K.~Horne\altaffilmark{23,R},
J.-R. Koo\altaffilmark{24},
D.~Kubas\altaffilmark{3},
J.-B.~Marquette\altaffilmark{3},
R.~Martin\altaffilmark{25},
J.W.~Menzies\altaffilmark{26},
K.C.~Sahu\altaffilmark{27},
J.~Wambsganss\altaffilmark{28},
A.~Williams\altaffilmark{25},
M.~Zub\altaffilmark{28}
\\ (The PLANET Collaboration) \\
J.Y.~Choi\altaffilmark{24},
D.L.~DePoy\altaffilmark{29},
Subo~Dong\altaffilmark{30},
B.S.~Gaudi\altaffilmark{2},
A.~Gould\altaffilmark{2},
C.~Han\altaffilmark{24},
C.B.~Henderson\altaffilmark{2},
D.~McGregor\altaffilmark{2},
C.-U.~Lee\altaffilmark{31},
R.W.~Pogge\altaffilmark{2},
I.-G.~Shin\altaffilmark{24},
J.C.~Yee\altaffilmark{2,32},
\\ (The $\mu$FUN Collaboration) \\
M.K.~Szyma{\'n}ski\altaffilmark{8},
J.~Skowron\altaffilmark{8},
R.~Poleski\altaffilmark{2,8},
S. Koz{\l}owski\altaffilmark{8},
{\L}.~Wyrzykowski\altaffilmark{8,33},
M.~Kubiak\altaffilmark{8},
P.~Pietrukowicz\altaffilmark{8},
G.~Pietrzy{\'n}ski\altaffilmark{8,34},
I.~Soszy{\'n}ski\altaffilmark{8},
K.~Ulaczyk\altaffilmark{8},
\\ (The OGLE Collaboration) \\ 
Y.~Tsapras\altaffilmark{35,36},
R.A.~Street\altaffilmark{35},
M. Dominik\altaffilmark{23,37},
D.M.~Bramich\altaffilmark{38,39},
P.~Browne\altaffilmark{23},
M.~Hundertmark\altaffilmark{23},
N.~Kains\altaffilmark{38},
C.~Snodgrass\altaffilmark{40},
I.A.~Steele\altaffilmark{41},
\\ (The RoboNet Collaboration) \\
I.~Dekany\altaffilmark{42},
O.A.~Gonzalez\altaffilmark{19},
D.~Heyrovsk\'y\altaffilmark{43},
R.~Kandori\altaffilmark{44},
E.~Kerins\altaffilmark{45},
P.W.~Lucas\altaffilmark{46},
D.~Minniti\altaffilmark{42},
T.~Nagayama\altaffilmark{44}, 
M.~Rejkuba\altaffilmark{19},
A.C.~Robin\altaffilmark{47}, and
R.~Saito\altaffilmark{42}
             } 
              

\altaffiltext{1}{Department of Physics,
    University of Notre Dame, 225 Nieuwland Science Hall, Notre Dame, IN 46556, USA; 
    Email: {\tt bennett@nd.edu}}
\altaffiltext{2}{Department of Astronomy, Ohio State University, 140 West 18th Avenue, Columbus, OH 43210, USA}
\altaffiltext{3}{UPMC-CNRS, UMR 7095, Institut d'Astrophysique de Paris, 98bis boulevard Arago, F-75014 Paris, France}
\altaffiltext{4}{Institute of Natural and Mathematical Sciences, Massey University, Auckland 0745, New Zealand}
\altaffiltext{5}{Department of Physics, Massachussets Institute of Technology, Cambridge, MA 02139, USA}
\altaffiltext{6}{Goddard Space Flight Center, Greenbelt, MD 20771, USA}
\altaffiltext{7}{Department of Earth and Space Science, Osaka University, Osaka 560-0043, Japan}
\altaffiltext{8}{Warsaw University Observatory, Al.~Ujazdowskie~4, 00-478~Warszawa, Poland}
\altaffiltext{9}{Technische Universit\"{a}t Wien, Wieder Hauptst. 8-10, A-1040 Vienna, Austria}
\altaffiltext{10}{Solar-Terrestrial Environment Laboratory, Nagoya University, Nagoya, 464-8601, Japan}
\altaffiltext{11}{Department of Physics, University of Auckland, Private Bag 92-019, Auckland 1001, New Zealand}
\altaffiltext{12}{Okayama Astrophysical Observatory, National Astronomical Observatory of Japan, Okayama 719-0232, Japan}
\altaffiltext{13}{Nagano National College of Technology, Nagano 381-8550, Japan}
\altaffiltext{14}{School of Chemical and Physical Sciences, Victoria University, Wellington, New Zealand}
\altaffiltext{15}{Tokyo Metropolitan College of Aeronautics, Tokyo 116-8523, Japan}
\altaffiltext{16}{Mt. John University Observatory, P.O. Box 56, Lake Tekapo 8770, New Zealand}
\altaffiltext{17}{University of Canterbury, Department of Physics and Astronomy, Christchurch 8020, New Zealand}
\altaffiltext{18}{IRAP, CNRS, Universit\'{e} de Toulouse, 31400 Toulouse, France}
\altaffiltext{19}{European Southern Observatory, Casilla 19001, Vitacura 19, Santiago, Chile}
\altaffiltext{20}{McDonald Observatory, Fort Davis, Texas 79734, USA}
\altaffiltext{21}{University of Tasmania, School of Mathematics and Physics, Hobart, TAS 7001, Australia}
\altaffiltext{22}{Department of Physics, University of Rijeka, Omladinska 14, 51000 Rijeka, Croatia}
\altaffiltext{23}{SUPA, University of St Andrews, School of Physics \& Astronomy, St Andrews, KY16 9SS, UK}
\altaffiltext{24}{Department of Physics, Chungbuk National University, Chongju 371-763, Korea}
\altaffiltext{25}{Perth Observatory, Walnut Road, Bickley, Perth 6076, WA, Australia}
\altaffiltext{26}{South African Astronomical Observatory, P.O. Box 9 Observatory 7925, South Africa}
\altaffiltext{27}{Space Telescope Science Institute, 3700 San Martin Drive, Baltimore, MD 21218, USA}
\altaffiltext{28}{Astronomisches Rechen-Institut, Zentrum f\"{u}r Astronomie der Universit\"{a}t Heidelberg (ZAH), 69120 Heidelberg, Germany}
\altaffiltext{29}{Department of Physics, Texas A\&M University, College Station, TX 77843-4242, USA}
\altaffiltext{30}{Kavli Institute for Astronomy and Astrophysics, Peking University, Hai Dian District, Beijing 100871, China}
\altaffiltext{31}{Korea Astronomy and Space Science Institute, Yuseong-gu 305-348 Daejeon, Korea}
\altaffiltext{32}{Sagan Fellow; Harvard-Smithsonian Center for Astrophysics, 60 Garden Street, Cambridge, MA 02138}
\altaffiltext{33}{Institute of Astronomy, University of Cambridge, Madingley Road, Cambridge CB3 0HA, UK}
\altaffiltext{34}{Universidad de Concepci\'on, Departamento de Astronom\'{\i}a, Casilla 160--C, Concepci\'on, Chile}
\altaffiltext{35}{Las Cumbres Observatory Global Telescope Network, 6740B Cortona Dr, Goleta, CA 93117, USA}
\altaffiltext{36}{School of Physics and Astronomy, Queen Mary University of London, Mile End Road, London, E1 4NS}
\altaffiltext{37}{Royal Society University Research Fellow}
\altaffiltext{38}{ESO Headquarters, Karl-Schwarzschild-Str. 2, 85748 Garching bei M\"{u}nchen, Germany}
\altaffiltext{39}{Qatar Environment and Energy Research Institute, Qatar Foundation, Doha, Qatar}
\altaffiltext{40}{Max Planck Institute for Solar System Research, 37191 Katlenburg-Lindau, Germany}
\altaffiltext{41}{Astrophysics Research Institute, Liverpool John Moores University, Liverpool CH41 1LD, UK}
\altaffiltext{42}{Pontificia Universidad Catolica de Chile, Casilla 306, Santiago 22, Chile}
\altaffiltext{43}{Institute of Theoretical Physics, Charles University, 18000 Prague, Czech Republic}
\altaffiltext{44}{Graduate School of Science, Nagoya University, Furo-cho, Chikusa-ku, Nagoya 464-8602, Japan}
\altaffiltext{45}{Jodrell Bank Centre for Astrophysics, University of Manchester, Oxford Road, Manchester, M13 9PL, UK}
\altaffiltext{46}{University of Hertfordshire, Hatfield, UK}
\altaffiltext{47}{Observatoire de Besancon Institut UTINAM, Universite Franche-Comte, CNRS-UMR 6213, BP 1615, 25010 Besancon Cedex, France}
\altaffiltext{M}{Microlensing Observations in Astrophysics (MOA) Collaboration}
\altaffiltext{P}{Probing Lensing Anomalies NETwork (PLANET) Collaboration}
\altaffiltext{U}{Microlensing Follow-up Network ($\mu$FUN Collaboration)}
\altaffiltext{O}{Optical Gravitational Lensing Experiment (OGLE) Collaboration}
\altaffiltext{R}{RoboNet Collaboration}



\begin{abstract}
We present the first microlensing candidate for a free-floating exoplanet-exomoon 
system, MOA-2011-BLG-262, with a primary lens mass of 
$M_{\rm host} \sim 4$ Jupiter masses hosting a sub-Earth mass moon. 
The data are well fit by this exomoon model, but an alternate 
star+planet model fits the data almost as well. Nevertheless, 
these results indicate the potential of microlensing to detect exomoons,
 albeit ones that are different from the giant planet moons in our 
 solar system. The argument for an exomoon hinges on the system being 
 relatively close to the Sun. The data constrain the product $M_L\pi_{\rm rel}$ where
$M_L$ is the lens system mass and $\pi_{\rm rel}$ is the lens-source relative
parallax. If the lens system is nearby (large $\pi_{\rm rel}$), then $M_L$ is 
small (a few Jupiter masses) and the companion is a sub-Earth-mass 
exomoon. The best-fit solution has a large lens-source relative 
proper motion, $\mu_{\rm rel} = 19.6 \pm 1.6\,$mas/yr, which would 
rule out a distant lens system unless the source star has an unusually 
high proper motion. However, data from the OGLE collaboration 
nearly rule out a high source proper motion, so the exoplanet+exomoon 
model is the favored interpretation for the best fit model. However, the 
alternate solution has a lower proper motion, which is compatible 
with a distant (so stellar) host. A Bayesian analysis does not favor 
the exoplanet+exomoon interpretation, so Occam's razor favors 
a lens system in the bulge with host and companion masses of 
$M_{\rm host} = 0.12{+0.19\atop -0.06}\,\msun$
and $m_{\rm comp} = 18{+28\atop -10}\,\mearth$, at a projected 
separation of $a_\perp = 0.84{+0.25\atop -0.14}\,$AU. The existence 
of this degeneracy is an unlucky accident, so current microlensing 
experiments are in principle sensitive to exomoons. In some 
circumstances, it will be possible to definitively establish the low 
mass of such lens systems through the microlensing parallax effect. 
Future experiments will be sensitive to less extreme exomoons.
\end{abstract}


\section{Introduction}
\label{sec-intro}

Gravitational microlensing occupies a unique niche among planet
detection methods \citep{bennett_rev,gaudi_araa}. 
While the radial velocity \citep{butler-catalog,mayor12} 
and transit methods \citep{borucki11,kepler_16mon} are most sensitive to planets 
in short period orbits, microlensing is most sensitive to planets orbiting beyond the
snow line \citep{mao91,gouldloeb92}
where the leading theory of planet formation, core accretion 
\citep{lissauer_araa,pollack96}, predicts that the most massive planets should
form. Ices, including water ice, can condense beyond the snow line
\citep{ida05,lecar_snowline,kennedy-searth,kennedy_snowline,thommes08},
and this means that the density of solid material in the proto-planetary
disk increases by a factor of a few beyond the snow line, so that
solid giant planet cores can form more rapidly. This is important
because the Hydrogen and Helium that comprise the vast majority of
the mass of gas giant planets is thought to be removed from 
proto-planetary disks in a few million years, so if gas giants do not
form relatively quickly, they cannot form at all. For this reason, theory suggests that 
it could be difficult for gas giants to form around M-dwarfs \citep{laughlin04}.

Radial velocity observations \citep{johnson07,johnson10} seemed to confirm
this picture, but microlensing observations \citep{gould10,cassan12} paint
a somewhat more complicated picture. Microlensing finds that
$17{+6\atop -9}$\% of stars have gas giant planets above 0.3 Jupiter
masses \citep{cassan12} with a host star sample strongly dominated by
M-dwarfs. However, the planets found by microlensing generally orbit
beyond the snow line, and the gas giants found by microlensing are often
low-mass gas giants with a mass similar to Saturn (at 0.3 Jupiter masses)
\citep{gould10}.
The combination of the relatively low-masses and wide orbits for the gas giants
found by microlensing along with their relatively wide orbits means that analogs
of most of the gas giant planets found by microlensing would not have been 
detected in the radial velocity surveys conducted to date. But a recent
combined analysis of precise radial velocity measurements and high-contrast 
imaging has been used to derive occurrence frequency of gas giants
orbiting M-dwarfs beyond the snow line \citep{montet13}, and they find that
their results are consistent with the microlensing results \citep{gould10,cassan12}.

Microlensing results also indicate that super-Earths or Neptunes are
substantially more common than gas giants beyond the 
snow line\citep{sumi10,cassan12}. This is roughly in line with the predictions of
the core accretion theory \citep{lissauer_araa,pollack96}, which predicts
that ``failed Jupiters" with masses of order $\sim 10\mearth$ should be 
quite common. These ``failed Jupiters" are expected to be much more 
common than gas-giants for M-dwarf stellar hosts \citep{laughlin04},
although some have argued that the gravitational
instability model could also make such planets \citep{boss06}.

Perhaps the most surprising microlensing result to date was the discovery
of a large population of planetary mass objects with no detectable host
star \citep{sumi11}. While some of these could be bound planets
in wide orbits \citep{quanz12}, the median separation of these planets is probably 
$>30\,$AU \citep{bennett12}, and it seems likely that many of them are unbound.
Although objects in the planetary mass range are now being found by direct
observation in the infrared \citep{delhome12,beichman13}, these infrared
surveys do not currently reach down to $\sim 1\,$Jupiter mass, where the
microlensing signal is seen.
Unbound planets are expected from a variety of processes, including
planet-planet scattering \citep{levison98,ford08,guillochon11}, star-planet scattering 
\citep{holman99,musielak05,doolin11,malmberg11,veras12,kaib13}, and
stellar mass loss and death \citep{veras11,veras_tout12,voyatzis13}.
However, most of these processes have been expected to generate fewer
of these unbound or very wide orbit planets than the 
$1.8{+1.7\atop -0.8}$ Jupiter-mass planets per main sequence star found
by microlensing \citep{sumi11}.

The evidence for this isolated planet population is statistical, and is based
on the distribution of the Einstein radius crossing times, $t_E$, for microlensing
events seen towards the Galactic bulge. But, these $t_E$ values depend
not only on the masses of the gravitational lenses, but also on the lens distances,
$D_L$, and their relative proper motions, $\mu_{\rm rel}$, with respect to
the source stars. For a subset of events, it is possible to obtain more
information. If the source star and lens object angular separation becomes
very much smaller than the angular Einstein radius, $\theta_E$, then finite
source effects may allow the measurement of the angular Einstein radius
\citep{macho-95b30}. In extreme cases, terrestrial microlensing parallax
effects can be measured, which will enable a direct measurement of the 
lens mass \citep{gould09,yee09} and a determination of the distance to the lens.

The relative proper motion, $\mu_{\rm rel}$, can also be measured for
binary lensing events, which often have caustic crossing features that
are infinitely sharp. Most binary events with planetary mass ratios of
$q \simlt 10^{-3}$ fall into this category. The event presented
in this paper is one such event, and it is also the shortest duration event 
detected to date with a
planetary mass ratio companion. The Einstein radius crossing time
is $t_E \simeq 3.8\,$days, and the mass ratio is $q\simeq 4.7\times 10^{-4}$.
While the photometry for this event is not precise enough to measure the
microlensing parallax effect, finite source effects are clearly measured.
This allows the source radius crossing time, $t_\ast$, to be measured,
and for the best fit model, this gives a lens-source relative proper motion
$\mu_{\rm rel} = \theta_\ast/t_\ast = 19.6 \pm 1.6\,$mas/yr, measured
in the inertial geocentric reference frame moving at the velocity
of the Earth at the time of the event.
Such a high relative proper motion
would usually imply that the lens must be nearby, particularly
since the source star proper motion is fairly tightly constrained by
the OGLE data \citep{skowron13}. This $t_\ast$ measurement
also leads to a determination of the angular Einstein radius
$\theta_E = t_E  \theta_\ast/t_\ast = 0.205\pm 0.016\,$mas, which
can be used in a mass-distance relationship. But, the Einstein
radius crossing time, $t_E$, is unusually small for an event
with such a small $t_*$ value, and this implies a small 
$\theta_E$ and a small lens mass of
only a few Jupiter masses. The large relative proper motion
implies that the lens is likely to be nearby, but the relatively small
$\theta_E$ value implies that the lens system mass must
be quite low if it is nearby. 

Unfortunately, the light curve for this event has one feature that
complicates the interpretation. The time interval between the
caustic entry and the caustic exit is quite similar to the source radius
crossing time, $t_\ast$, and as a result, the $\sim 1.4\,$hr caustic
crossing feature may be fit by models in which the sum of the 
source diameter crossing time and the caustic entry-exit interval
is $\sim 1.4\,$hrs. We find two distinct, nearly degenerate solutions,
which follow this caustic feature duration condition. The second
best solution is disfavored by $\Delta\chi^2 = 2.9$ and has a 
$t_\ast$ value that is larger than the value for the best fit solution
by a factor of 1.68. Since the remaining parameters, except for the
lens separation, are nearly the same, this implies a 
geocentric relative proper
motion of $\mu_{\rm rel} = 11.6 \pm 0.9\,$mas/yr. This value is
still much larger than average, and it is larger than the relative
proper motion reported for any other planetary mass ratio
binary microlensing event. This high relative proper motion 
also suggests a nearby lens. However,
as we shall see below in Section~\ref{sec-anal}, this preference for
nearby lenses is not enough to overcome the preference for stellar-mass
lens primaries due to their larger Einstein radii. So, when the appropriate
prior on the mass function is included, a stellar mass or brown dwarf
host is preferred. 

This paper is organized as follows. We discuss the data and its
collection in Section~\ref{sec-data}, and in Section~\ref{sec-lc}
we present our light curve models. Section~\ref{sec-radius}
presents the calibration of the optical data and the source radius
estimate, while Section~\ref{sec-source-con} describes 
constraints on the source proper motion using OGLE data and the analysis
of high resolution infrared adaptive optics (AO) observations made
with the Keck-2 telescope. In Section~\ref{sec-anal}, we present
a Bayesian analysis, which relates the observed light curve
parameters to the physical parameters of the lens system,
and in Section~\ref{sec-tpar}, we discuss how very high
cadence observations from multiple sites could 
resolve the ambiguities in similar events using the
terrestrial parallax effect.
Finally in Section~\ref{sec-conclude}, we present our
conclusions.

\section{Data and Photometry}
\label{sec-data}

Microlensing event MOA-2011-BLG-262 
[$({\rm RA},{\rm DEC})=(18^{\rm h}\ 00^{\rm m}\ 23.48^{\rm s}, 
-31^\circ \ 14'\ 42.93'')$ and $(l,b)=(-0\rlap.^\circ 3693,-3\rlap.^\circ 9245)$]
was identified by the MOA alert system \citep{bond01} and announced by the
MOA group at 2011 June 26 14:16 UT, based on microlensing survey 
data from the MOA-II telescope at Mt. John University Observatory (MJUO) in
New Zealand. It was immediately recognized by the
MOA, PLANET and $\mu$FUN groups that this was a short duration, high magnification
event that would be highly sensitive to planetary mass ratio companions
\citep{griest98,rhie00}. The $\mu$FUN group sent an email alert announcing
this fact.
Follow-up observations were immediately begun by the PLANET Collaboration,
using the 1.0m telescope at Canopus Observatory in Tasmania, and by MOA using the 
0.61m Boller \& Chivens telescope, also at MJUO. The observing
cadence on the MOA-II 1.8m telescope was increased from one observation
every 50 minutes to one observation every 2 minutes, and then cut back
to one observation every 7 minutes.
These follow-up data were enabled by the rapid identification of the event
by the MOA Alert system \citep{bond01}, which identified the event based
on the first three observations of the night and allowed the alert to 
be used $\sim 50$ minutes after the third observation of the night.
Because of the short duration of the event, it was unclear whether observatories
in Chile would be able to get useful color information for the source star,
so the MOA group obtained both $V$ and $I$-band photometry from the
0.61m Boller \& Chivens telescope.

At about the time that these follow-up observations began, the
leading limb of the source star crossed a caustic due to a planetary
mass ratio companion to the primary lens object. This caustic
feature was observed by the MOA-II 1.8m, the MJUO 0.61m,
and the Canopus 1.0m telescopes until well after the trailing limb of the 
source exited the caustic $\sim 1.4\,$hrs later. A peak magnification
of $A_{\rm max} = 75$ was reached, as shown in Figure~\ref{fig-lc}.
The observations from New Zealand and Tasmania continued
until night ended, when the magnification had dropped to $A < 30$.
During the final hour of observations from New Zealand and
Eastern Australia, the Robonet Collaboration was able to
observe with the 2.0m Faulkes South Telescope (FTS) located at
the Siding Springs Observatory, but unfortunately,
this was after the planetary anomaly had ended.

Observations from Chile began at about ${\rm HJD} = 2455739.5$,
when the magnification was about $A\approx 10$. These observations
were made by the OGLE group using the 1.3m Warsaw 
University telescope and the OGLE-IV camera at the Las Campanas
Observatory and the $\mu$FUN group using the 1.3m SMARTS telescope
at the Cerro Tololo InterAmerican Observatory. Because the event
was no longer at high magnification when it could be observed
from Chile, the $\mu$FUN group focused on obtaining multi-color data
in the $V$, $I$, and $H$ passbands. While the CTIO $V$ and $I$-band
data constrained the source color with less S/N than the $V$ and
$I$-band data from the MJUO 0.61m, the $H$-band data were unique,
and proved to be necessary to normalize follow-up high
resolution AO data. The CTIO $V$ and $I$ data were also
used help determine the dust extinction in the vicinity of the target star.
While the $\mu$FUN
observations were made as a result of the MOA alert of this event,
OGLE had been observing this event previously as a part of the
OGLE-IV survey. The event was discovered independently by the
OGLE Early Warning System \citep{ogle-ews} and announced as OGLE-2011-BLG-0703.
Because the lens is located in a
lower cadence OGLE-IV field, the alert was triggered exactly on the peak
night and distributed just after the peak magnification.

We also obtained high angular resolution infrared AO follow-up data from the
Keck-2 telscope (as discussed in Section~\ref{sec-source-con}, and 
wide-field infrared images from the InfraRed Survey Facility (IRSF) at
the South African Astronomical Observatory (SAAO) to help calibrate
the Keck data. However, data from the Vista Variables in the Via Lactea (VVV)
project \citep{minniti-vvv} became available, and this was used instead of the
IRSF data.

The light curve data were reduced to photometry using a number of
implementations of the difference imaging photometry method
\citep{tom96,ala98}. The MOA-II telescope data, consisting
of 4884 observations in the MOA custom red band (roughly
equivalent to Cousins $R$+$I$), and the MJUO 0.61m
data were reduced with the MOA-pipeline \citep{bond01}.
The MJUO data consist of 143 $V$-band and 168 $I$-band
measurements. The 59 $I$-band observations from Canopus
Observatory were reduced with pySIS \citep{albrow09}; the
37 FTS SDSS-$I$-band observations were reduced
with the Robonet pipeline \citep{bramich08}, and the  298
OGLE-IV $I$-band observations were reduced with the OGLE
pipeline \citep{ogle-pipeline}.

As with most microlensing events, the error bars calculated by
these photometry codes give only a rough estimate of the actual
photometry errors. These are sufficient to find the best fit models,
but error bars that give $\chi^2/{\rm dof} \approx 1$ are needed in
order to estimate the errors on the physical parameters of the lens
system. We have therefore followed the standard procedure of 
adding 0.3\% in quadrature to each error estimate, and then
renormalized the error bars to give  $\chi^2/{\rm dof} = 1$ for
a preliminary planetary model (a version of the $s<1$ fast model
in this case.). Experience with the analysis of a large
number of other events indicates that the final results have
no significant dependence on the details of this procedure or
on which preliminary model is used to determine the error bar
renormalization factors.

We computed the source-star limb-darkening coefficients from Kurucz's 
ATLAS9 stellar model atmosphere grid 
\citep{kurucz93a,kurucz93b,kurucz94} using the method described in 
\citet{heyrovsky07}. For the photometric band of each light curve we 
used instrument-specific response functions obtained by combining the 
respective filter transmission and CCD quantum efficiency curves. We 
interpolated the limb-darkening coefficients for stellar effective 
temperature $T_{\rm eff}= 5520\,K$, surface gravity $\log\,g= 4.1$, 
microturbulent velocity $v_{\rm t}= 2\,$km/sec, and solar metallicity 
from the values obtained for the Kurucz-model parameter grid.
These parameters were selected to be consistent with the source
star properties presented in Section~\ref{sec-radius}.
The computed linear limb-darkening coefficients for the four light 
curves with sufficient caustic-crossing points are 0.5291, 0.4861, 
0.6587, and 0.4605 for MOA red, Canopus $I$, MJUO $V$, and MJUO $I$, 
respectively.

\section{Light Curve Models}
\label{sec-lc}

\begin{figure}
\plotone{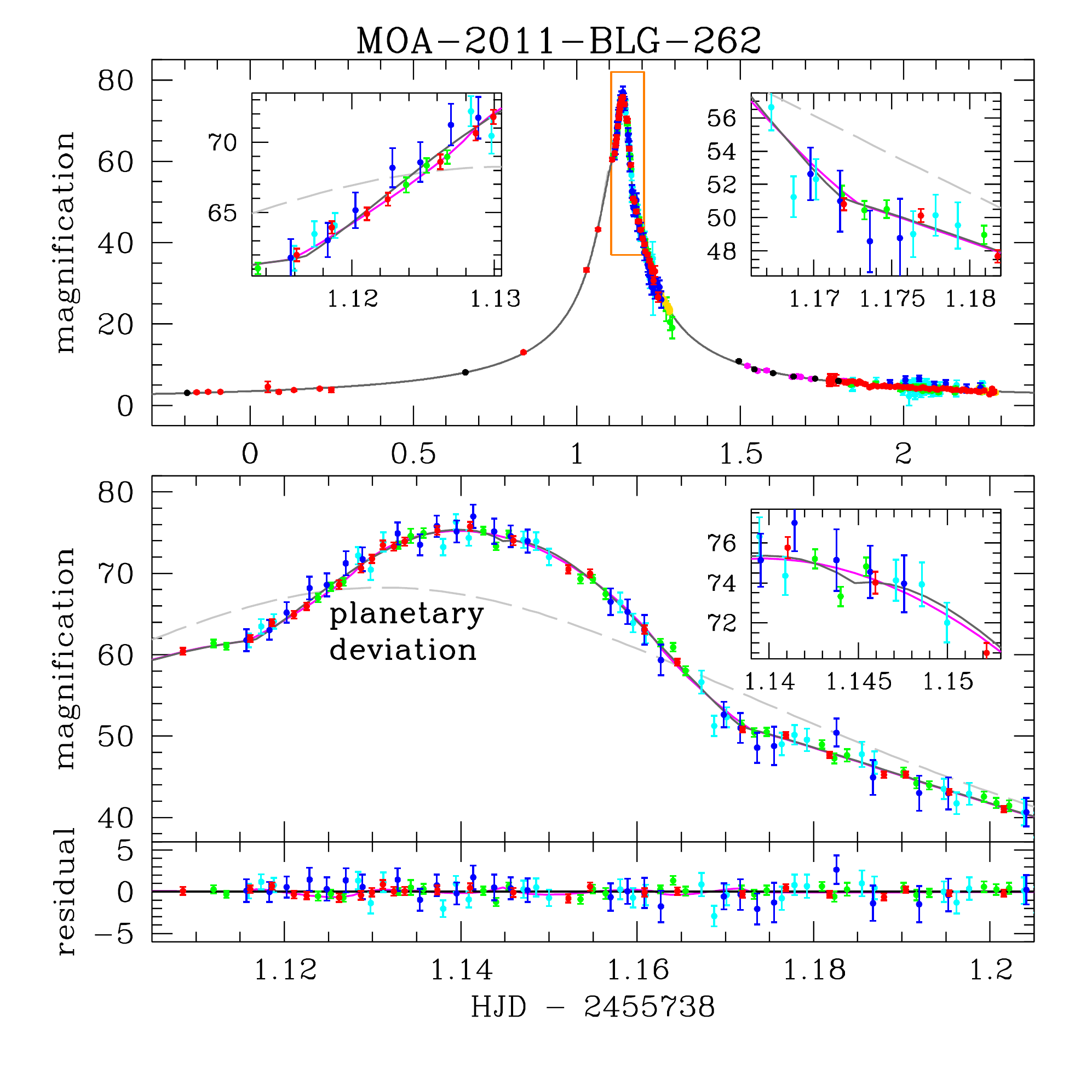}
\caption{The light curve of event MOA-2011-BLG-262 with data from
the MOA 1.8m (red), Mt. John University Observatory $I$ and $V$-bands
(green and cyan), Canopus $I$-band (blue), CTIO $I$-band (magenta),
OGLE $I$-band (black), and Faulkes Telescope South (gold). The best fit 
model is indicated by the black curve,
and the magenta curve indicates an alternative planetary model,
which almost fits the light curve except for the limb crossings shown
in the inserts. This model has a $\chi^2$ larger by $\Delta\chi^2 = 2.91$
when compared to the best fit model.
\label{fig-lc}}
\end{figure}

\begin{figure}
\plotone{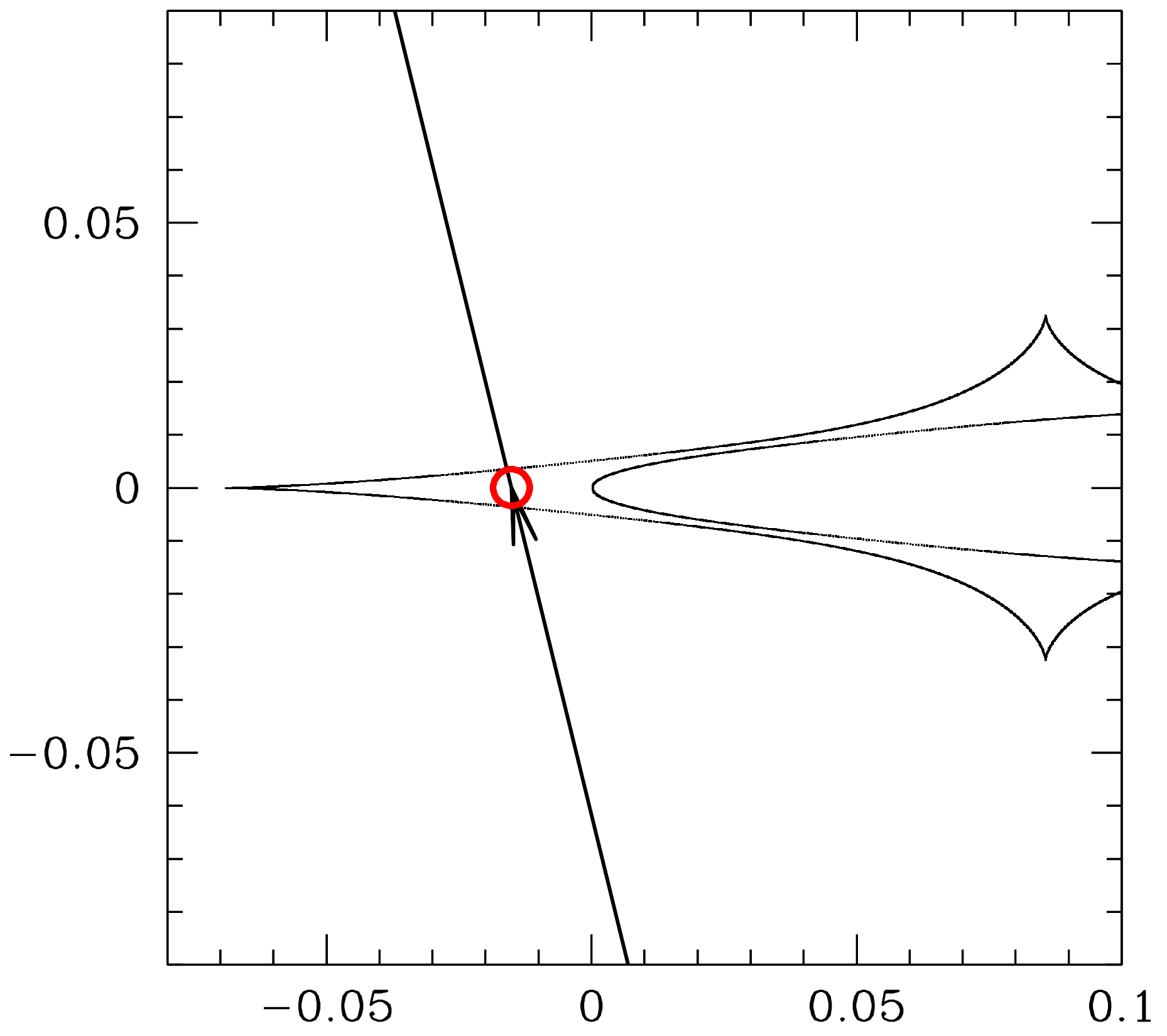}
\caption{ The caustic geometry for the best fit MOA-2011-BLG-262 ``fast" model.
The red circle indicates the source star size.
\label{fig-caustic_fast}}
\end{figure}

The light curve of this event, shown in Figure~\ref{fig-lc}, was modeled using the
image centered ray-shooting method \citep{bennett96,bennett-himag}. The
global fit strategy used was the initial condition grid search method of
\citet{bennett-himag}. However, after this grid search was run, we also
did a grid search in the lens separation, $s$, in order to ensure that
all degenerate solutions were recovered. We find four distinct 
$\chi^2$ minima, and the parameters of these solutions are given
in Table~\ref{tab-mparams}. These four solutions are due to two
degeneracies. The first degeneracy is the well known \citep{griest98}
2-fold $s \leftrightarrow 1/s$ degeneracy that occurs for most, but not 
all \citep{gaudi-ogle109,bennett-ogle109,miyake11}, planets found in high magnification 
events. The two best fit models, labeled ``fast" in Table~\ref{tab-mparams}
have nearly identical parameters, except that the best model has
$s = 0.9578$, and its $s> 1$ counterpart has $s = 1.0605$ and a $\chi^2$ 
that is larger by $\Delta\chi^2 = 0.64$. These values of $s$ do not 
precisely correspond a $s \leftrightarrow 1/s$ degeneracy since
$1/0.9578 = 1.0441 \neq 1.0605$.  This is because 
the shape of the central caustic is perturbed by the proximity of the
planetary caustic for $s \approx 1$, and in this case, at $s = 0.9578$, the
planetary and central caustics have merged to form a so-called
``resonant caustic\rlap," as shown in Figure~\ref{fig-caustic_fast}.

The other degeneracy is more unusual, and as alluded to in Section~\ref{sec-intro},
it occurs because the caustic entrance-to-exit time interval is similar to the 
source radius crossing time, $t_\ast$. More typically, the source radius
crossing time is much smaller than the caustic entrance-to-exit interval
\citep{bond04,miyake11,bachelet12} or else the caustic is much narrower
than the source \citep{ogle390,dong-moa400,janczak10}. There are a variety
of solutions with different caustic widths and different $t_\ast$ values,
such that the caustic entrance-to-exit time interval is the same. The light
curve features that distinguish these light curves are subtle and not easily
resolved, particularly in the present case with a caustic entrance-to-exit time interval
of only 80 minutes.
We find two distinct solutions for both the $s<1$ and $s>1$ cases. The alternate, 
``slow\rlap," models have $s$ values that are further from unity than the fast model, but
as Table~\ref{tab-mparams} indicates, the other parameters, except for $t_\ast$,  are 
quite similar to those of the fast models. In particular, the best fit source brightnesses 
and color differ by $<0.01\,$mag for the fast and slow models. As discussed in
Section~\ref{sec-radius}, this implies that the angular sources star radius,
$\theta_\ast$, values will be nearly identical for these two models. Thus,
the main physical difference in the two models is in the (geocentric) relative proper
motion, $\mu_{\rm rel} = \theta_\ast/t_\ast$. The source radius crossing times
are $t_\ast = 0.01316$ and $t_\ast = 0.02217$ for the $s < 1$ fast and slow
models, respectively. In Section \ref{sec-radius},
we derive $\theta_\ast$ and find that $\mu_{\rm rel} = 19.6 \pm 1.6\,$mas/yr
for the fast models and $\mu_{\rm rel} = 11.6 \pm 0.9\,$mas/yr for the 
slow models. (This is the reason for the ``fast" and ``slow" designations.)

Figure~\ref{fig-lc} shows the event light curve with both of the best fit
$s < 1$ models. The black curve is the best fit fast, $s < 1$ model, which
is the overall best fit, and the magenta curve is the best fit slow model
(also with $s < 1$), which is disfavored by $\Delta\chi^2 = 2.91$.
The parameters of these models are given in the third and fifth columns of
Table~\ref{tab-mparams}. These models are difficult to distinguish, but they
do differ slightly at the caustic entry and exit, as well as at the midpoint of the
caustic feature where the leading limb of the source begins to exit the
caustic just as the trailing limb enters for the fast model. This gives an abrupt change in
slope, although the amplitude of the change is small. These three regions
are highlighted in the three insets in Figure~\ref{fig-lc}. The MJUO 0.61m
and Canopus data are not precise enough to clearly distinguish these models.
The MOA-2 1.8m telescope data might be precise enough to distinguish these 
models if the density of observations were much higher (say one observation
every minute), but with the actual observing cadence of one observation every
$\sim 5$ minutes, these two models cannot be definitively distinguished.

\section{Calibration and Source Radius}
\label{sec-radius}

\begin{figure}
\plotone{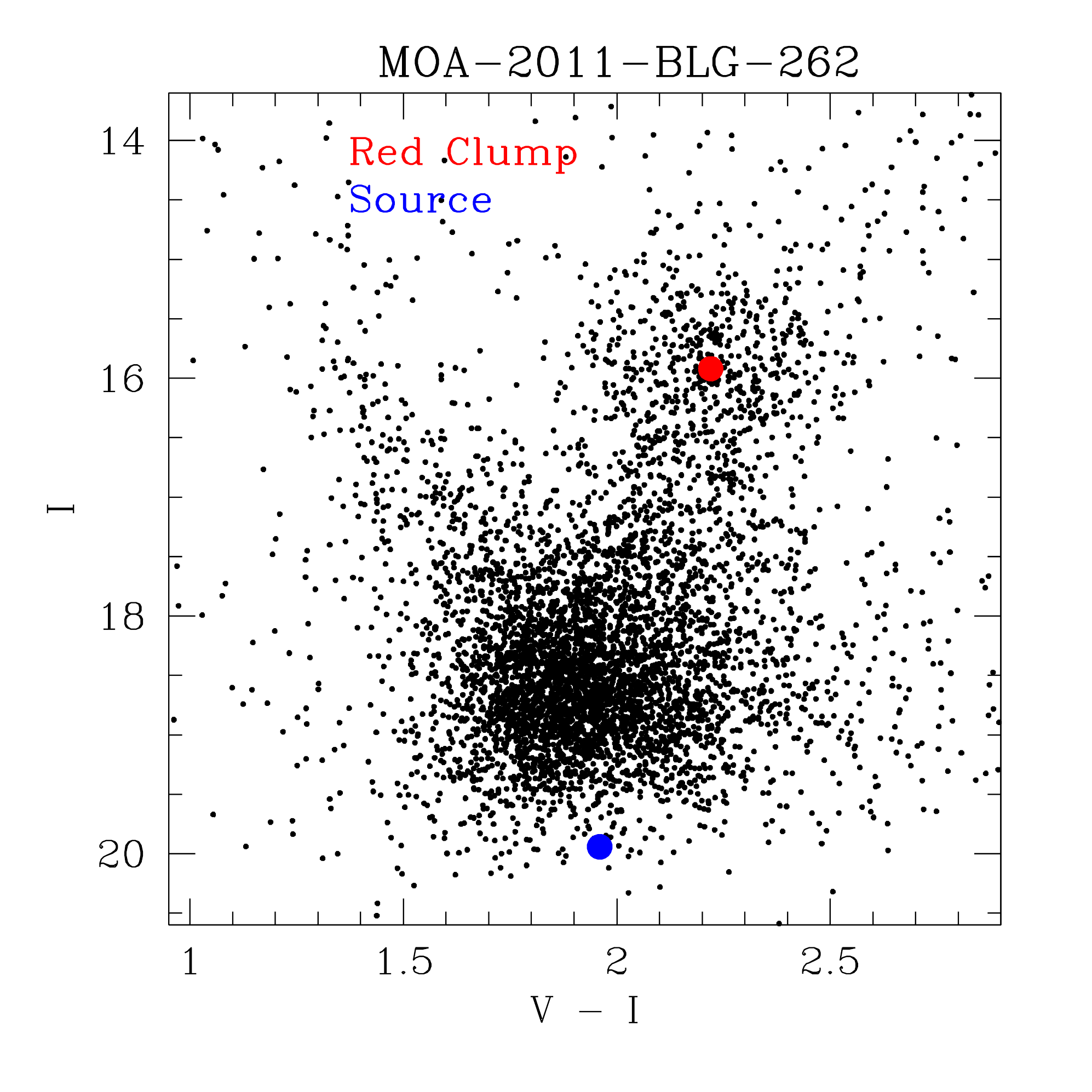}
\caption{The color magnitude diagram (CMD) of the stars in the CTIO images,
calibrated to the OGLE-III catlog
within $135^{\prime\prime}$ of MOA-2011-BLG-262. The red spot indicates
red clump giant centroid, and the blue spot indicates the source magnitude
and color.
\label{fig-cmd}}
\end{figure}

In order to measure the angular Einstein radius, $\theta_E = \theta_\ast t_E/t_\ast$, 
we must determine $\theta_\ast$ from the dereddened brightness and color of the 
source star \citep{kervella08}. For most microlensing events colors are obtained
from $V$ and $I$-band measurements from either OGLE or CTIO, but this
event was so short that it could only be observed at high magnification from
New Zealand or Australia. Therefore, we obtained $V$ and $I$-band photometry
from the MJUO 0.61m B\&C telescope, as shown in Figure~\ref{fig-lc}.
These data were calibrated using stars in the OGLE-III catalog 
\citep{ogle3-phot} within 2 arc minutes of the source star. However, this
target is located near CCD edge in the OGLE-III reference frame, and the
OGLE-III catalog does not report both $V$ and $I$ magnitudes for
stars within 199 pixels (or $51\rlap.^{\prime\prime} 7$)
of the field edge, including the immediate vicinity
of the source star. This means that an estimate of the extinction in the 
foreground of the source would be especially susceptible to errors
due to differential reddening.
To avoid this possibility, we use the CTIO $V$ and $I$-band photometry
of this CCD to generate the color magnitude diagram (CMD), shown in
Figure~\ref{fig-cmd}. The CTIO data have been calibrated to the OGLE-III
data base in the same way as the MJUO 0.61m B\&C data.
The centroid of the red clump giant feature in this CMD is 
$I_{\rm cl} = 16.04$, $V_{\rm cl} = 18.22$, and
$ (V-I)_{\rm cl} = 2.18$. We chose to use the CTIO data instead of
the MJUO 0.61m data for the CMD
in order to minimize blending effects, since the CTIO images have
significantly better seeing. But the red clump centroid from the
MJUO 0.61m CMD is within 0.02 mag of the CTIO red clump centroid
in both $(V-I)$ and $I$, so this choice does not affect our results.

\citet{nataf13} find that the unreddened red clump
magnitude and color at this Galactic position ($l = -0.3693$, $b = -3.9245$) are
$I_{\rm cl0} = 14.47$ and $(V-I)_{\rm cl0}= 1.06$, and this implies that
$A_I = 1.57$ and $A_V = 2.93$. This yields an extinction law with
$R_{VI} = 2.40$ and $R_v(BV) = 2.69$, assuming the \citet{cardelli89} extinction law. 
We can then use the best fit source magnitudes listed in Table~\ref{tab-mparams}
to determine the dereddened source magnitudes. For the best fit
(fast, $s < 1$) model, we find $I_{s0} = 18.359$ and $(V-I)_{s0} = 0.840$,
while for the best fit slow model they are almost identical,
$I_{s0} = 18.367$ and $(V-I)_{s0} = 0.841$. With these dereddened magnitudes,
we can use \citet{kervella08} to give the angular source radius. This
yields $\theta_* = 0.778 \pm 0.059\,\mu$as for the fast ($s<1$) model
and $\theta_* = 0.776 \pm 0.059\,\mu$as for the slow ($s<1$) model.

Of course, the main difference between the fast and slow models is their
different source radius crossing times: $t_* = 0.01316\,$days for the
fast model, and $t_* = 0.02217\,$days for the slow model (both with
$s < 1$). These source radius crossing times give angular 
Einstein radii of $\theta_E = \theta_* t_E/t_* = 0.205\pm 0.015\,$mas
and $\theta_E = 0.122\pm 0.009\,$mas for the fast and slow models,
respectively. The implied lens-source relative geocentric proper motion values
are $\mu_{\rm rel} = \theta_*/t_* = 19.6\pm 1.6\,$mas/yr for the fast model
and $\mu_{\rm rel} = 11.6\pm 0.9\,$mas/yr for the slow model.

\section{Source Star Constraints}
\label{sec-source-con}

\subsection{Source Proper Motion from OGLE Data}
\label{sec-source-pm}

With the ground-based, seeing limited data used by microlensing surveys, it is generally
not possible to measure the proper motion of stars as faint as 
MOA-2011-BLG-262S at $I_s \simeq 19.9$. The central Galactic bulge fields,
where most microlensing events are observed, are very crowded. They are so
crowded that the seeing disks generally have more than one star with $I \simlt 20$.
Thus, PSF fitting photometry codes,
like DoPHOT \citep{dophot} or DAOPHOT \citep{allframe} are usually unable to identify
individual main sequence stars. The faint star-like images seen in these images are generally
blends of multiple stars, so their apparent proper motion would actually be
some average of the proper motion of the multiple stars contributing to
each blend. As a result, attempts to measure proper motions with microlensing
survey data \citep{sumi04,rattenbury07} are limited to relatively bright stars.
High angular resolution Hubble Space Telescope (HST) \citep{koz06} data is
needed to measure the proper motion of the faint main sequence and turn-off
stars in the bulge.

The situation is significantly improved for faint microlensing event source
stars with good quality light curve measurements. The microlensing signal
allows both the source star position and brightness
to be determined from the difference images, and microlensing
model. This additional information allows its proper motion to be 
measured using a method developed by \citet{skowron13}. They apply
their method to MOA-2011-BLG-262S, and find its proper motion to be
$\muSbold_{,\rm hel} = (-2.3,-0.9)\pm (2.8,2.6)\,$mas/yr in a (North,East) Galactic 
Heliocentric coordinate frame.

In order to compare with the geocentric relative proper motion values,
$\mu_{\rm rel}$, determined from the light curve models, we must 
convert this source proper motion value to the inertial geocentric
frame that moves at the velocity that the Earth had at the light
curve peak. This velocity was $(v_N,v_E) = (-0.2,29.2)\,$km/sec.
Assuming a source distance of $8.3\,$kpc, this implies a
source proper motion of $\muSbold_{,\rm geo} = (-2.3,-1.7)\pm (2.8,2.6)\,$mas/yr
in the geocentric frame appropriate for this event. These are the values
used for the Bayesian analysis presented in Section~\ref{sec-anal}.

\subsection{Keck Adaptive Optics Observations}
\label{sec-keck}

\begin{figure}
\includegraphics[width=3.5in]{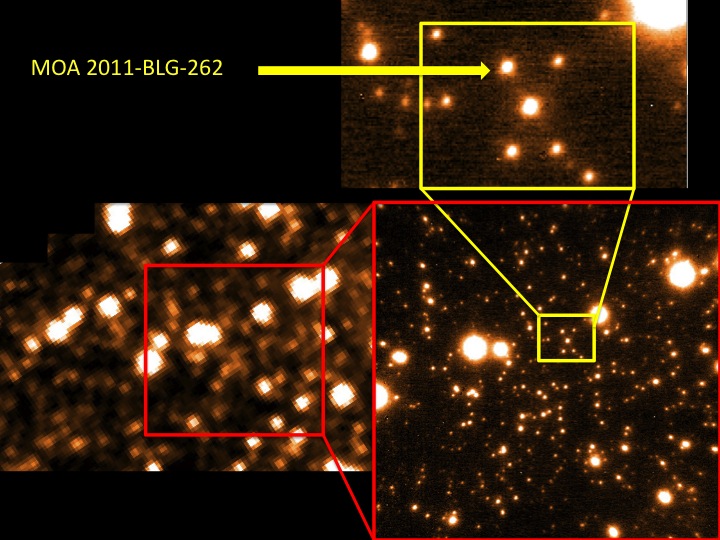}
\includegraphics[width=2.2in]{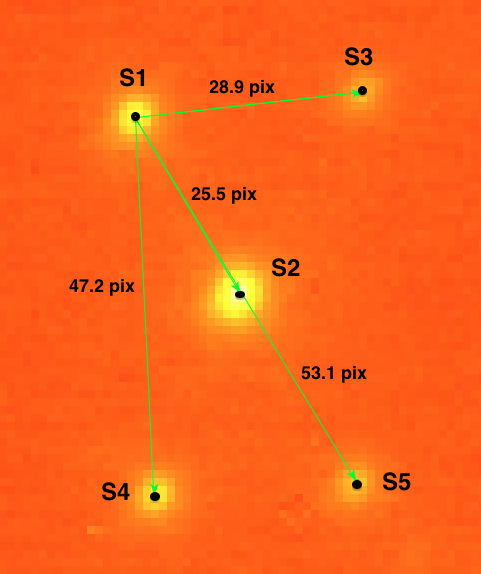}
\caption{The left panel shows a $K$-band image from the VISTA 4m telescope
from the VVV survey. The field observed by the Keck-2 telescope 
in $K$ and a zoom of this field are shown in the middle panel. The arrow indicates 
the microlensing source star, and it is 
separated by 0.51 arcsec from its nearest neighbor. The right panel is
shows a Keck-2 AO close up of the source star, S1, and 4 neighboring stars.
The distances between the target and its neighbors are given in 
$0.02^{\prime\prime}$ pixels.
\label{fig-keck_im}}
\end{figure}

$J$, $H$ and $K$ adaptive optics (AO) images of MOA-2011-BLG-262 were taken with
the NIRC2 instrument on the Keck-2 telescope on May 6, 2012, using the
Laser Guide Star system. This was long after the peak of the event, so the 
source was no longer magnified. The NIRC2 medium field camera was used
with a field of view of $20''\times 20''$, which has a 0.02 arcsec pixel size.
(Due to $2"$ dithering, stacked images have a size of $24''\times 24''$.) 
Figure~\ref{fig-keck_im} shows the $K$-band image stack (of $20\times 30\,$sec 
exposures) compared to a VVV image \citep{minniti-vvv} in the same band taken 
by the VISTA 4m wide field infrared telescope at Paranal. The high resolution 
Keck images reveal five stars that were not resolved in the photometric light curve
data used for modeling, and the star labeled S1 in Figure~\ref{fig-keck_im} is 
identified as the source star through an astrometric comparison of difference images 
taken near peak magnification \citep{skowron13}. The Keck AO images were taken less
than a year after the microlensing peak, and this implies that the lens-source
separation at the time of the images is $\simlt 20\,$mas. Since the seeing in the
Keck images is $0.10^{\prime\prime}$, a stellar lens would have its image unresolved from the source
star. While this field is crowded with stars 
when observed in seeing-limited ground-based images, the Keck AO image appears 
relatively uncrowded, with the vast majority of the stars well isolated.
In the following, we consider S1 as the source+lens position. A hypothetical 
unrelated star would be unresolved from, and add flux to the measured S1 flux,
if its separation from the target is $<200\,$mas.

The $J$, $H$ and $K$ Keck magnitudes have been measured with PSF photometry 
and calibrated to the 
2MASS \citep{2mass_cal}\footnote{Improved calibrations are available at \\
{\tt http://www.ipac.caltech.edu/2mass/releases/allsky/doc/sec6\_4b.html}}
system using the $JHK$ images from the VVV survey, as an intermediary 
between 2MASS and Keck. The calibrated magnitudes of star S1,
which would include both the source and the lens are
\begin{equation}
(J,H,K)_{\rm S1}=(18.64,18.15,18.15) \pm (0.10,0.07,0.10) 
\label{eq-magS1}
\end{equation}
The 5 stars S1-S5 are not resolved in the VVV images, but stars S4 and S5 are
far enough away that they are more likely to contribute to the background than
the detected source in the VVV image. The detected flux in $H$ band at the 
lens+source position is $H_{\rm S,VVV}=17.03 \pm 0.12$, which is nearly 
the same to the Keck measurement of the two brightest stars S1 and S2,
 $H_{\rm S1+S2,Keck}=16.95\pm0.10$.

\subsection{Lens Flux Upper Limits from Keck Observations}            

We have $H$-band data taken with the CTIO 1.3m SMARTS during the 
microlensing event that allow us to determine the magnitude of the source 
in this passband. The best fit source magnitudes are
$H_{S,{\rm fast}}= 18.220 \pm 0.040$ and $H_{S,{\rm slow}}= 18.226 \pm 0.040$.
These are to be subtracted from the total S1 target flux, which must include
both the source and the lens system, $H_{\rm S1} = 18.15 \pm 0.07$. These
measurements differ by less than 1-$\sigma$, there is no significant detection
of flux from the lens. But they do allow a lens flux of
$H_{L,{\rm fast}} \leq 21.16 \pm 1.10$ and $H_{L,{\rm slow}} \leq 21.07 \pm 1.00$,
where we use an inequality because of the possibility of additional flux from
an unrelated star.
There error bars on these source brightnesses were determined using a
linear approximation, which is actually not valid, as the uncertainties 
do not follow Gaussian statistics (even approximately) when expressed in
magnitudes. (The error bars are valid 1-$\sigma$ uncertainties when converted
into flux units using a linear approximation, however.)
The 2-$\sigma$ upper limits on the lens brightness for the
fast and slow models are $H_{L,{\rm fast}} > 19.85$ and $H_{L,{\rm slow}} > 19.83$.

\begin{figure}
\plottwo{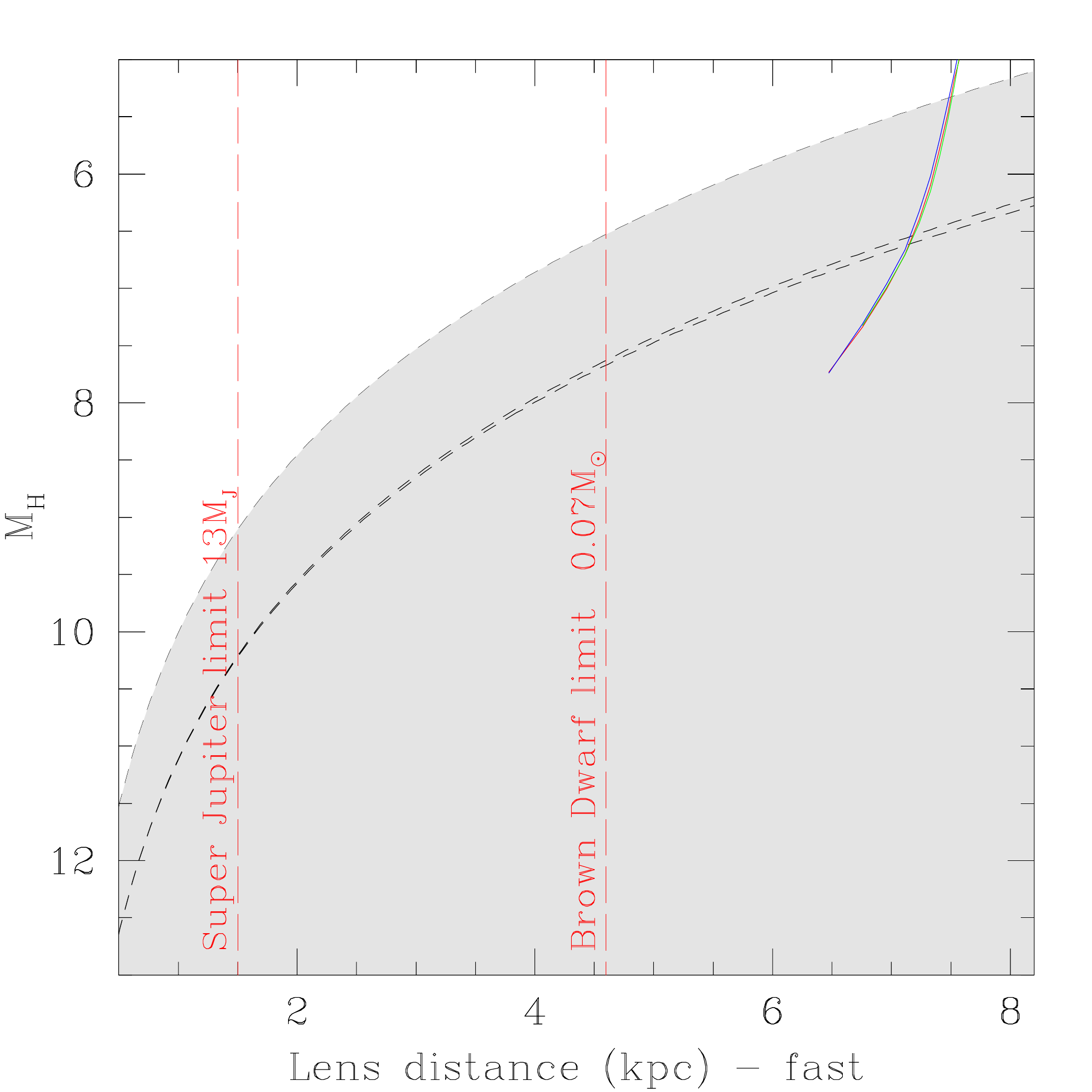}{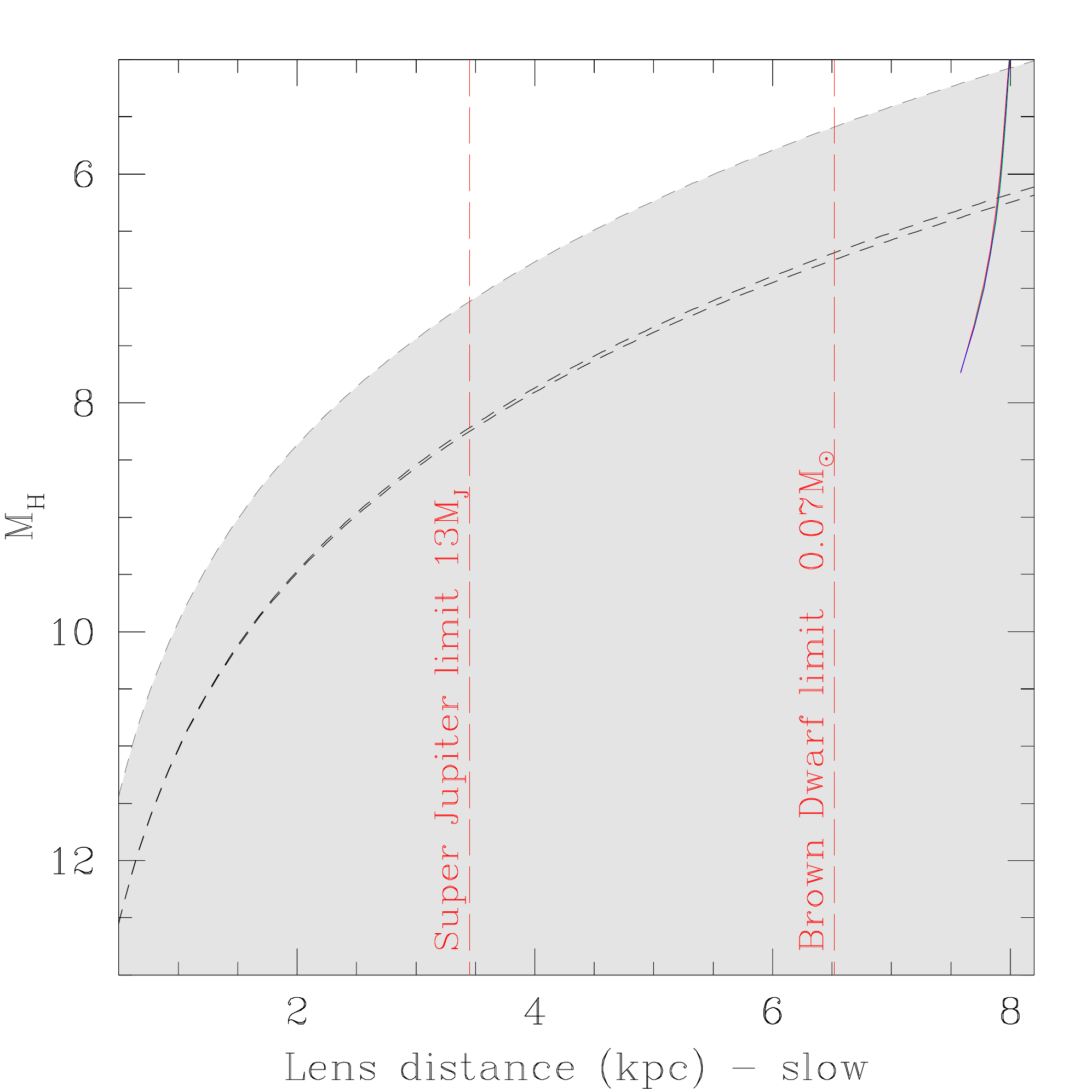}
\caption{Upper limits on the lens flux measured by Keck AO observations in H-band. 
The allowed region is shaded. The dashed black curves show the best fit magnitude distance relations 
for two extinction laws. The red dashed lines show the super Jupiter and brown dwarf 
ranges. The colored curves are some isochrones for main sequence stars from \citep{an07}.
\label{fig-keck_magD}}
\end{figure}

The lens magnitude estimates and limits can be expressed in terms
of the absolute magnitude as
$m_{\rm H}=H_L-A_H-5\log (D_L/10\rm pc)$,
where $A_H$ is the extinction. From \citep{vvv_extinct},
we obtain an $H$-band  total extinction of 
$A_{H,{\rm Car}}=0.39 \pm 0.11$ \citep{cardelli89} or 
$A_{H,{\rm Nish}}=0.315 \pm 0.09$ \citep{nish09}. 
Assuming a linear expression for the extinction along the line of sight, we 
show the upper limits of the lens flux in H band in Figure~\ref{fig-keck_magD}.
These plots assume a fixed source distance of $D_S = 8.0\,$kpc.
The left and right panels indicate the relations for the fast and slow models.
The magnitude distance relations for the two extinction laws are given by the
nearly overlapping black dashed curves in each plot, and the grey dashed lines and
shading indicate the extent of the 2-$\sigma$ allowed region of parameter space.
As a reference, we draw in red dashed lines the ranges for super Jupiters 
and brown dwarfs. We have also included isochrones for main sequence stars 
\citep{an07} in the upper right of each plot.  We take the oldest population from 
\citet{an07} of 4 Myrs and the following range for metallicity: 
$0.0 \leq [F_e/H] \leq +0.3$ ($[F_e/H]=0.0$ (dark blue), $[F_e/H]=0.2$ (green), 
and, $[F_e/H]=0.3$ (red)). The allowed ranges for stellar 
hosts are on the lower part of these isochrones, below the Keck curves. 
The crossing points between the distance-magnitude and isochrone curves are the 
following for the two different models $D_L=7$ kpc, and $D_L=7.7$ kpc
for the fast and slow models, respectively. These imply lens host masses
of $M_{\rm host} =0.36\msun$ and $M_{\rm host} = 0.41\msun$, for the fast and slow
models, respectively. The implied transverse velocities with respect to the
source, which is known to be moving relatively slowly \citep{skowron13} are
quite large. For the fast model, this velocity is $V_{\rm rel}=677\,$km/s, which is
likely above the escape velocity, while for the slow model, it is 
$V_{\rm rel}=442$, which is very high for a bulge star.

If the fast model is correct, then as we will see below in Section~\ref{sec-anal},
a host mass of only a few Jupiter masses is favored (with our
assumed mass function prior). In this
case it is sensible to ask the question of whether there is a host star
with a separation too wide to be found by microlensing \citep{quanz12}.
The distance to the planet+moon system favored by the fast model
is $D_L \sim 500\,$pc. Since a star at the bottom on the main sequence
has an absolute $H$-band magnitude of $M_H \approx 10.0$, its
brightness at $500\,$pc would be $H < 18.8$ even if virtually all the
dust extinction is in the foreground. Thus, we can rule such a host 
star blended with the lens, and the only possibility for a host star would be if
it was one of the resolved stars in the Keck images. However, all the
nearby stars have $(J-K)\simlt 1$, which excludes them as possible
host stars because if they were at $\sim 500\,$pc, they would have to 
be very low-mass, very red stars.

\section{Bayesian Analysis}
\label{sec-anal}

Because we are unable to measure a microlensing parallax signal for this
event, we have only two light curve parameters, $t_E$ and $\theta_*$ that
can be used to constrain the lens mass, $M_L$, distance, $D_L$, and 
lens-source relative transverse velocity, $v_\perp$. With 2 parameters to
constrain 3 unknowns, we have a 1-parameter family of solutions, or
a mass-distance relation,
\begin{equation}
M_L = {c^2\over 4G} \theta_E^2 {D_S D_L\over D_S - D_L} 
       =  {c^2\over 4G} \theta_E^2 {{\rm AU}\over \pi_{\rm rel}}
       = 0.9823\,\msun \left({\theta_E\over 1\,{\rm mas}}\right)^2\left({x\over 1-x}\right)
       \left({D_S\over 8\,{\rm kpc}}\right) \ ,
\label{eq-m_thetaE}
\end{equation}
where $x=D_L/D_S$ and $\theta_E= \theta_* t_E/t_*$, as discussed in 
Section~\ref{sec-radius}. The lens-source relative parallax is given
by $\pi_{\rm rel} = {\rm AU}\left(D_S^{-1}-D_L^{-1}\right)$.

The relative proper motion, $\mu_{\rm rel}$, provides a strong constraint
on the properties of the lens system. While the Galactic disk moves at about
the same velocity as the Sun, due to the flat Galactic rotation curve, the bulge
has an average velocity of zero. So, the average proper motion of a source
in the bulge is just given by the Galactic rotation speed divided by the distance
to the Galactic center. This gives $\mu \simeq 6.4\,$mas/yr for the
mean proper motion of a bulge star. A 1-dimensional velocity dispersion of 
$\sim 100\,$km/sec at the distance of the Galactic center gives a
1-dimensional proper motion dispersion of $\sigma_\mu \simeq 2.6\,$mas/yr.
The velocity dispersion in the disk is about 3 times smaller than in the bulge,
but the proper motion is inversely proportional to the distance, so the $\sigma_\mu$
values will be similar for a lens one third of the way to the Galactic
center, at $D_L \simeq 2.7\,$kpc. 

We can use these numbers to estimate the expected relative proper
motion values for bulge and disk lenses. If the lens, as well as the source,
is in the bulge, then the expected $\mu_{\rm rel}$ is given by the
quadrature sum of the 1-dimensional dispersions in the Galactic
$l$ and $b$ directions \citep{koz06} for both the lens and 
source\footnote{To get this result, we count events by the probability
that an event is in progress at a given time, rather than by the event
rate. This accounts for the fact that the detection efficiency for
a planetary microlensing signal is approximate proportional to
the event duration.}, or 
$\mu_{\rm rel} \simeq 2\sigma_\mu \simeq 5.2\,$mas/yr. For a disk
lens at $D_L \simeq 2.7\,$kpc, the average proper motion dispersion
will be about the same, but there will also be the mean proper motion
difference of  $\mu \simeq 6.4\,$mas/yr between the disk and bulge.
So, the average relative proper motion
will be the quadrature sum of the mean proper motion and the dispersion,
which yields a typical lens-source relative proper motion value of
$\mu_{\rm rel} \simeq 8.2\,$mas/yr.
Since the relative proper motion for both our fast and slow solutions
is larger than this, the $\mu_{\rm rel}$ values for both solutions
favor a disk lens at $D_L < 2.7\,$kpc. However, the degree by which
a bulge lens is disfavored is quite different. For the fast solution,
the measured proper motion value, of $\mu_{\rm rel} = 19.6\,$mas/yr 
is $3.5\times$ larger than the dispersion in the proper motion difference
between two bulge stars, $\mu \simeq 5.6\,$mas/yr. For the slow
solution $\mu_{\rm rel} = 11.6\,{\rm mas/yr} = 2.1 \times$ the
bulge-bulge $\mu_{\rm rel}$ dispersion. 
The Gaussian probabilities of a bulge-bulge relative proper motion outlier
with $\mu_{\rm rel}$ given by the values for our fast and slow solutions
are $0.0022$ and 0.11, respectively. So, the fast solution
very strongly favors a nearby lens, while the slow solution appears
to only slightly disfavor a bulge lens system. (In principle, the
OGLE measurement of source proper motion could change this
argument, but the error bars on the source proper motion are nearly
equal to the measured bulge proper motion dispersion, so the OGLE
measurement has little effect at this level of analysis.)

Another important prior assumption for the Bayesian analysis is the
mass function for the host mass. The only study that has looked
at the mass function for the full range of masses from stellar down to
planetary masses is that of \citet{sumi11}. This study didn't have much
leverage on the mass function below about a Jupiter mass, so most 
of the models consider only a delta-function mass function for the 
planetary-mass part of the mass function. A much better choice for
this analysis is the power-law planetary mass function presented
in the Supplementary Information (SI) section of \citet{sumi11}. This
is shown in Figure~S11 of that paper, and the parameters of
this power law model are given Table~S3. This model includes
stellar remnants and has a broken power law of the form
$dN/d\log M=M^{1-\alpha}$, with 
$\alpha = 2.0$ for $0.70 \leq M/\msun \leq 1.00$,
$\alpha = 1.3$ for $0.08 \leq M/\msun \leq 0.70$,
$\alpha = 0.49$ for $0.01 \leq M/\msun \leq 0.08$, and
$\alpha = 1.3$ for $10^{-5} \leq M/\msun \leq 0.01$.
For $\alpha = 1$, we have an equal number of objects
per logarithmic mass interval, and this mass function averages
close to $\alpha = 1$ over the interval from low
mass stars to planetary mass objects. The ratios of main sequence
stars to brown dwarfs to planets are 1:0.73:5.5 with this mass
function, but most of the planetary mass objects have masses much less
than that of Jupiter, because the planetary part of the mass function
extends over three decades in mass. If we restrict ourselves to 
planetary mass objects with masses of order a Jupiter 
mass, then the numbers of main sequence stars, brown dwarfs, and planets
are similar with this mass function. 

While our assumed mass function implies that the densities of
potential stellar, brown dwarf, and planetary mass are similar, this
does not mean that the lensing probabilities are similar for these
three populations. This is because the Einstein ring radius,
$R_E = \sqrt{(4GM/c^2) D_L(D_S-D_L)/D_S}$, is proportional
to $\sqrt{M}$. Since the typical star has $\sim 400\times$ Jupiter's
mass, the lensing rate due to stars is enhanced by a factor
of $\sim 20$ with respect to lensing by planetary mass objects.
The star density in the
Galactic bulge is a factor of $\sim 5$ larger than the local
density in the disk. All told, this implies a factor of $\sim 100$
preference for stellar lenses over planetary mass lenses, so
we should expect that the slow solution will not favor planetary
mass lenses.

We use a Galactic model including a barred bulge, a spheroid,
and thin and thick disks, using the functional forms for the
stellar densities from \citet{robin03}, but we use truncated Gaussians
for the velocity distributions. We do not allow stars with velocities that
exceed the assumed escape velocity cutoff of $550\,$km/sec in the 
thin and thick disks, as well as in the
spheroid. The assumed bulge escape velocity is $600\,$km/sec. 
Because of the high relative proper motion implied by the models
for this event, the results of the Bayesian analysis may depend
on the assumed velocity distributions of the Galactic model,
so we have normalized the bulge velocity dispersions to the
Hubble Space Telescope (HST) proper motion measurements 
of \citet{koz06}. Using a bar rotation velocity of 50 km/sec/kpc, 
we can match the proper motion of the 5 \citet{koz06} fields
within $< 1^\circ\llap .5$ of MOA-2011-BLG-262 with 
velocity dispersions of $103.8\,$km/s and $96.4\,$km/s in the
Galactic longitude and latitude directions, respectively. 

The Bayesian priors discussed so far are all based on measurements
of the properties of the stars, brown dwarfs and
planets that are part of the Galaxy. But, we still must make one
prior assumption for which we have no data to guide us. We will
assume that probability of a lens primary to host a companion with
the measured mass ratio at a separation of $\sim R_E$ is independent
of the mass of the host. For stellar mass hosts, we have data
that show that planetary mass ratio secondaries are common at these
separations \citep{gould10,cassan12}, but for brown dwarf and planetary
mass hosts, this is simply an assumption.

The Bayesian analysis is done with a collection of Markov Chains
centered at the parameter space locations of the 4 solutions listed
in Table~\ref{tab-mparams}. These are the fast and slow solutions for
both $s<1$ and $s>1$. The Markov chains for each local minima
are weighted by the $\chi^2$ difference between the best fit
model in each region of parameter space. There were two Markov Chains
for each of these models, and each chain had about 350,000 links.
The result of this
Bayesian analysis is given in Figure~\ref{fig-lens_prop}, while
Figures~\ref{fig-lens_prop_fast} and \ref{fig-lens_prop_slow}, respectively, give the results
for the fast and slow models only.

Figure~\ref{fig-lens_prop} shows the probability distribution of the
host and planet masses, their projected separation, and the
distance to the lens system, under the assumption that planetary
and stellar mass hosts are equally likely to host a moon
or planet with the observed mass ratio.
There are two peaks in each distribution - one centered on
planetary mass hosts at $D_L = 0.64{+0.32\atop -0.21}\,$kpc, and the
other centered on stars just above the Hydrogen-burning
threshold in the central Galactic bulge at $D_L = 7.0{+0.9\atop -1.0}\,$kpc.
The probability drops to virtually zero at $D_L \sim 3.6\,$kpc, so we have
two separate distributions: the nearby, planetary mass hosts and
the stellar/brown dwarf hosts in the bulge. For the planetary mass
distribution, the host and companion masses are 
$M_{\rm host} = 3.6{+2.0\atop -1.7}\,M_{\rm Jup}$ and
$m_{\rm comp} = 0.54{+0.30\atop -0.19}\,\mearth$, respectively, and their
projected separation is $a_\perp = 0.13{+0.06\atop 0.04}\,$AU.
For the stellar/brown dwarf host solutions, the host and companion masses are
$M_{\rm host} = 0.12{+0.19\atop -0.06}\,\msun$ and
$m_{\rm comp} = 18{+28\atop -10}\,\mearth$, at a 
projected separation of $a_\perp = 0.84{+0.25\atop -0.14}\,$AU.

Figures~\ref{fig-lens_prop_fast} and \ref{fig-lens_prop_slow} show that
the planetary mass hosts come almost exclusively from 
the fast solution Markov Chains and that the slow solutions predict
a stellar mass host located in the bulge.

We should also note that without the OGLE constraint on the 
source proper motion \citep{skowron13}, the Bayesian probability distribution for the
fast solutions would resemble Figure~\ref{fig-lens_prop} with the
similar sized
peaks for both nearby disk planetary mass hosts and 
bulge low-mass star hosts. The OGLE source proper motion constraint
nearly rules out the possibility that the source is in the high velocity
tail of the bulge velocity distribution. For the slow solution, it
is still possible to recover the observed lens source relative
proper motion of $\mu_{\rm rel} = 11.6 \pm 0.9\,$mas/yr if the 
lens is in the high velocity tail of the bulge velocity
distribution. And the fact that stars have a much higher lensing
cross section than planetary mass options enables these solutions
to be favored over the nearby planetary mass solutions even though
the $\mu_{\rm rel}$ value itself does favor a nearby disk lens.

The situation with the fast solutions is somewhat different. The very high
relative proper motions, $\mu_{\rm rel} = 19.6 \pm 1.6\,$mas/yr, for these 
models would generally require that both the lens and the source
belong to the high velocity tail of the bulge velocity dispersion. Since the
OGLE source proper motion constraint nearly rules this out, bulge lenses are
significantly disfavored with these models. So, the fast solutions, if one of them
is correct, do seem to suggest that the lens system could consist
of a rogue planet of a few Jupiter masses orbited by a sub-Earth-mass
moon. If the fast model was the only light curve model consistent with
the light curve, then the planet+moon interpretation could be confirmed
with a higher precision measurement of the source proper motion using
the Hubble Space Telescope or adaptive optics observations.

\section{Terrestrial Parallax}
\label{sec-tpar}

The simultaneous observations of the MOA-2011-BLG-262 microlensing
event from MJUO on New Zealand's South Island,
and Canopus in Tasmania, Australia are potentially very useful. Although these observatories
are only separated by about $2000\,$km, this separation is potentially 
large enough so that the observed light curves would be significantly different
as observed from the two observatories, due to the terrestrial parallax effect
\citep{gould09,yee09}. According to the Bayesian analysis presented in Section~\ref{sec-anal},
the favored solution for the fast model has a host mass of $M_{\rm host} = 3.6{+2.0\atop -1.7}\,M_{\rm Jup}$
at a distance of $D_L = 0.64{+0.32\atop -0.21}\,$kpc. 
Such a model predicts terrestrial parallax magnification
differences between MJUO and Canopus of $\sim 0.5$\%. This compares to
error bars of $\sim 0.7$\% for the MOA-II data at the peak, so if the Canopus
error bars were comparable (and preferably with a higher observing cadence), terrestrial
parallax could be measured for the case that the fast planetary-mass host model is correct.
Unfortunately, due to a telescope hardware issue, the images taken from Canopus 
during this event had unusually poor image quality, so the Canopus photometry was
not precise enough for a terrestrial parallax measurement even if the 
fast light curve model with a nearby, planetary mass lens system was correct.

If we had photometry that allowed a terrestrial microlensing parallax measurement,
then the implications of the nearly degenerate fast and slow solutions would be 
relatively modest. The lens masses implied by the two models would differ by 
only the factor of 1.7 uncertainty in $t_\ast$. In fact, the uncertainty could be
much less as better photometry might be able to determine which of these
models is correct. As we saw in
Section~\ref{sec-anal}, the Bayesian analysis that we must use without a parallax
measurement results in a large uncertainty in the properties of
the lens system.

Fortunately, it should be possible to definitely distinguish similar models
for future events
if they are observed with high cadence from multiple sites. Very high
cadence observations on 1-2m class telescopes are able to measure the
light curves precisely enough to distinguish similar models
\citep{gould06}, such as the
fast and slow models for MOA-2011-BLG-262. If these high cadence observations
are taken from observatories separated by thousands of kilometers,
then the terrestrial parallax effect can be used to measure the lens
masses \citep{gould_yee_terpar13}. Future wide-field, multi-site surveys
\citep{kmtnet,kmtnet_pipe} will significantly increase the microlensing exoplanet detection
rate. However, real-time planetary signal detection, such as that demonstrated by
the MOA Collaboration for this event, and high cadence
follow-up observations will be necessary to take full advantage of
these new powerful surveys. The larger telescopes of the older
microlensing follow-up groups, such as PLANET \citep{ogle390} and
$\mu$FUN \citep{gould10} can contribute some of the required
high cadence follow-up observations, but the development of large
robotic networks of 1m class telescopes, such as the Las Cumbres
Robotic Telescope Network \citep{lcogt}, will substantially improve
the rate of terrestrial microlensing parallax mass measurements.
Thus, if systems resembling the planetary-mass host models for 
MOA-2011-BLG-262 are common, the combination of high cadence 
microlensing surveys, rapid realtime event detection by these surveys, 
and high cadence follow-up observations should enable the definitive
discovery of rogue exoplanets with moons of nearly an Earth mass
within a few years.

\section{Summary and Conclusions}
\label{sec-conclude}

We have presented the analysis of microlensing event
MOA-2011-BLG-262, which is the shortest duration microlensing
event with a planetary mass ratio of $q = 4.7\times 10^{-4}$. The best fit 
``fast" model implies a high lens-source relative proper of 
$\mu_{\rm rel} = 19.6\pm 1.6\,$mas/yr, which favors a lens at
a distance of order $D_L \sim 500\,$pc. The lens mass-distance
relation (see equation~\ref{eq-m_thetaE}) that comes from
this $\mu_{\rm rel}$ measurement implies that the lens primary
for this model would have mass of 3 or 4 Jupiter masses, and
the planetary mass ratio companion would be a moon
with a mass of $< 1\mearth$. However, the Bayesian analysis
reveals that a stellar lens host is not excluded by this fast
model. Also, the source radius crossing time, $t_*$, is similar
to the time interval between the caustic entry and exit, and this allows
for another solution is a somewhat larger $t*$ value, which implies a
smaller lens-source relative proper motion,
$\mu_{\rm rel} = 11.6\pm 0.9\,$mas/yr. This ``slow" solution is disfavored
by only $\Delta\chi^2 = 2.91$, so it definitely cannot be excluded.
Although the $\mu_{\rm rel}$ for this event is also high enough to
favor a nearby disk lens, it is much more compatible with a fast moving 
stellar lens in the bulge than the fast solution.

\begin{figure}
\plotone{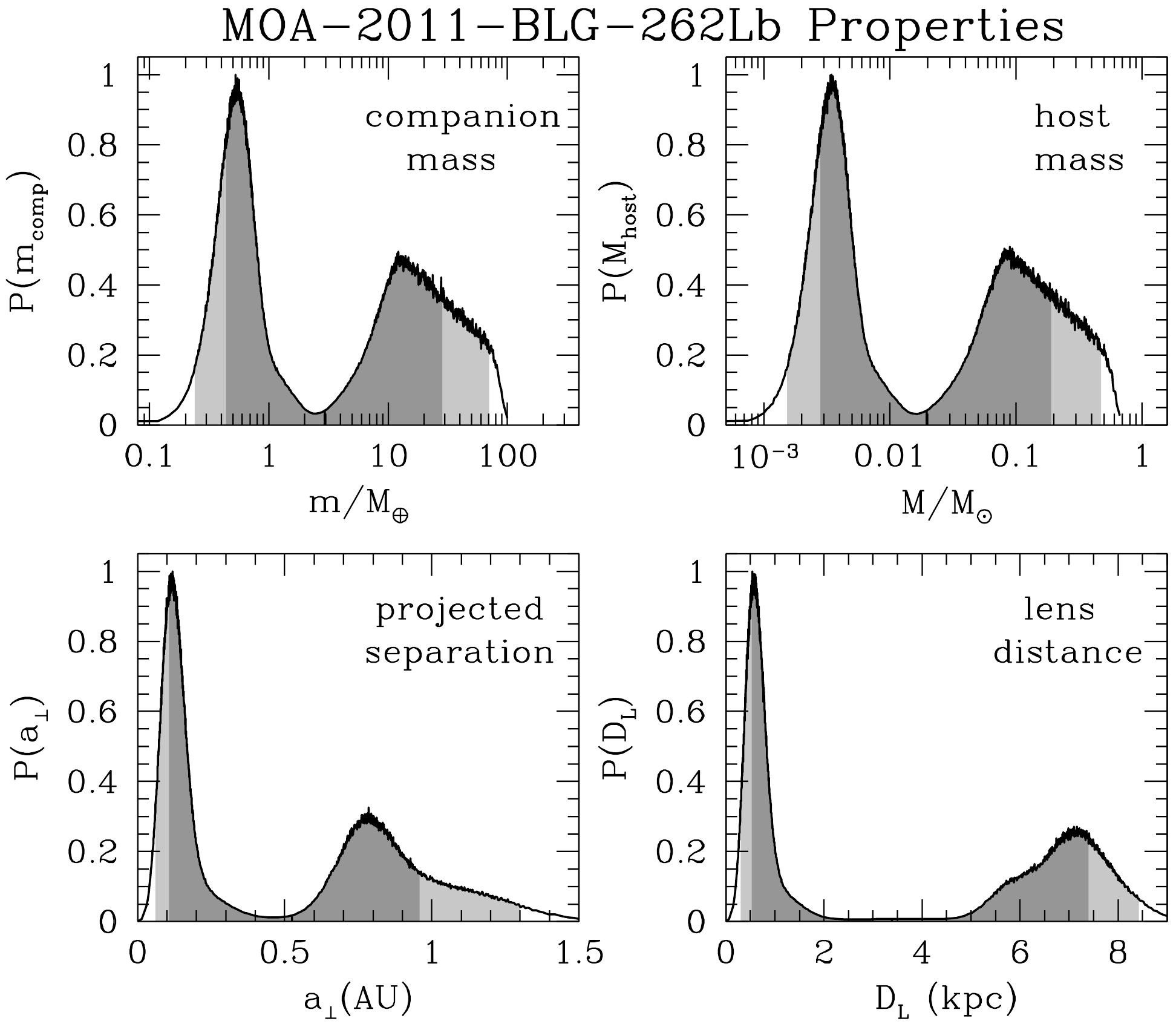}
\caption{Lens system properties from our Bayesian analysis
using a standard Galactic model under the assumption
that primary lenses of all masses have an equal probability of 
hosting a planetary mass ratio companion. This includes both
the wide and close versions of the fast and slow solutions weighted
by Galactic model priors and the best fit $\chi^2$ for in each
region of parameter space.
\label{fig-lens_prop}}
\end{figure}

These relatively large relative proper motion values can largely
be attributed to the motion of the lens, because the OGLE 
Collaboration has used a novel method to show that the source
proper motion is low \citep{skowron13}. In order to determine the
likely physical properties of the lens system, we perform a
Bayesian analysis with a standard Galactic model with 
bulge velocities normalized to the proper motion measurements
of \citet{koz06}. Since planetary mass hosts are possible
for this event, we have used the power-law planetary mass model
of \citet{sumi11} (see the Supplementary Information) as our
mass function prior. The only prior that is unconstrained by data
is the probability that a lens primary of a given mass will host
a companion lens with a mass ratio of $q = 4.7\times 10^{-4}$
at a separation of about an Einstein radius. For simplicity, we
assume that this probability is independent of the primary
lens mass. With such an assumption, we find two distinct
types of lens systems that could explain this event, and
with our assumed prior probabilities, the probability of these
two solutions are roughly equal, as indicated 
in Figure~\ref{fig-lens_prop}.

\begin{figure}
\plotone{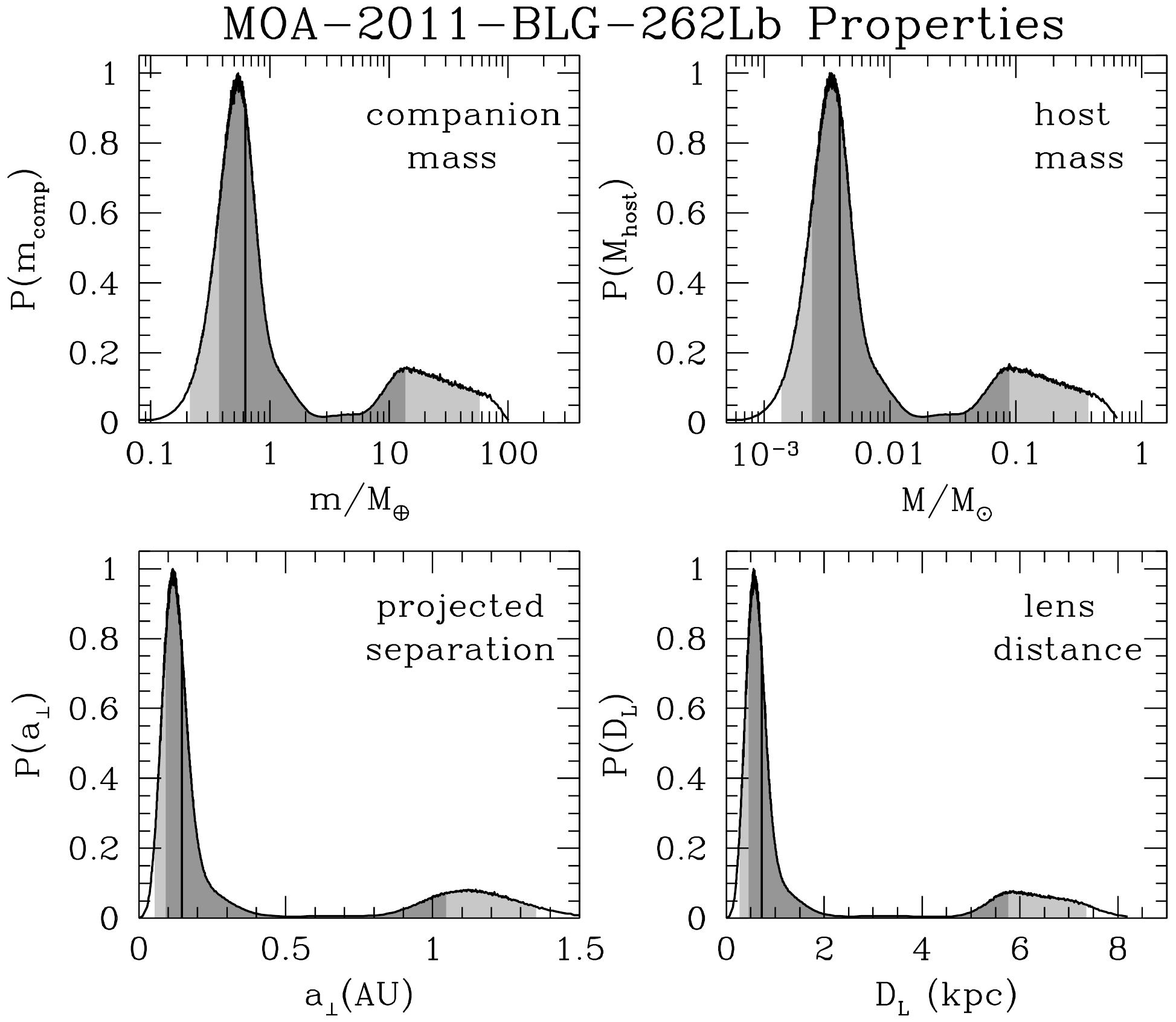}
\caption{Lens system properties for the wide and close versions of
the ``fast" solution only. This
solution clearly favors a host mass of a few Jupiter masses orbited
by a companion of $\sim 0.5\mearth$. The lens system would be
$\simlt 600\,$pc from the Earth. 
\label{fig-lens_prop_fast}}
\end{figure}

The fast solution favors, but does not require, a planetary mass 
for the host, as shown in Figure~\ref{fig-lens_prop_fast}. However,
high angular resolution observations with adaptive optics or
the Hubble Space Telescope could improve the precision of
the proper motion measurement \citep{skowron13}, and
exclude the non-planetary mass host models for the fast solution.
So, it it is only the existence of the slow solution that removes
the possibility of confirming the planet+moon model.

\begin{figure}
\plotone{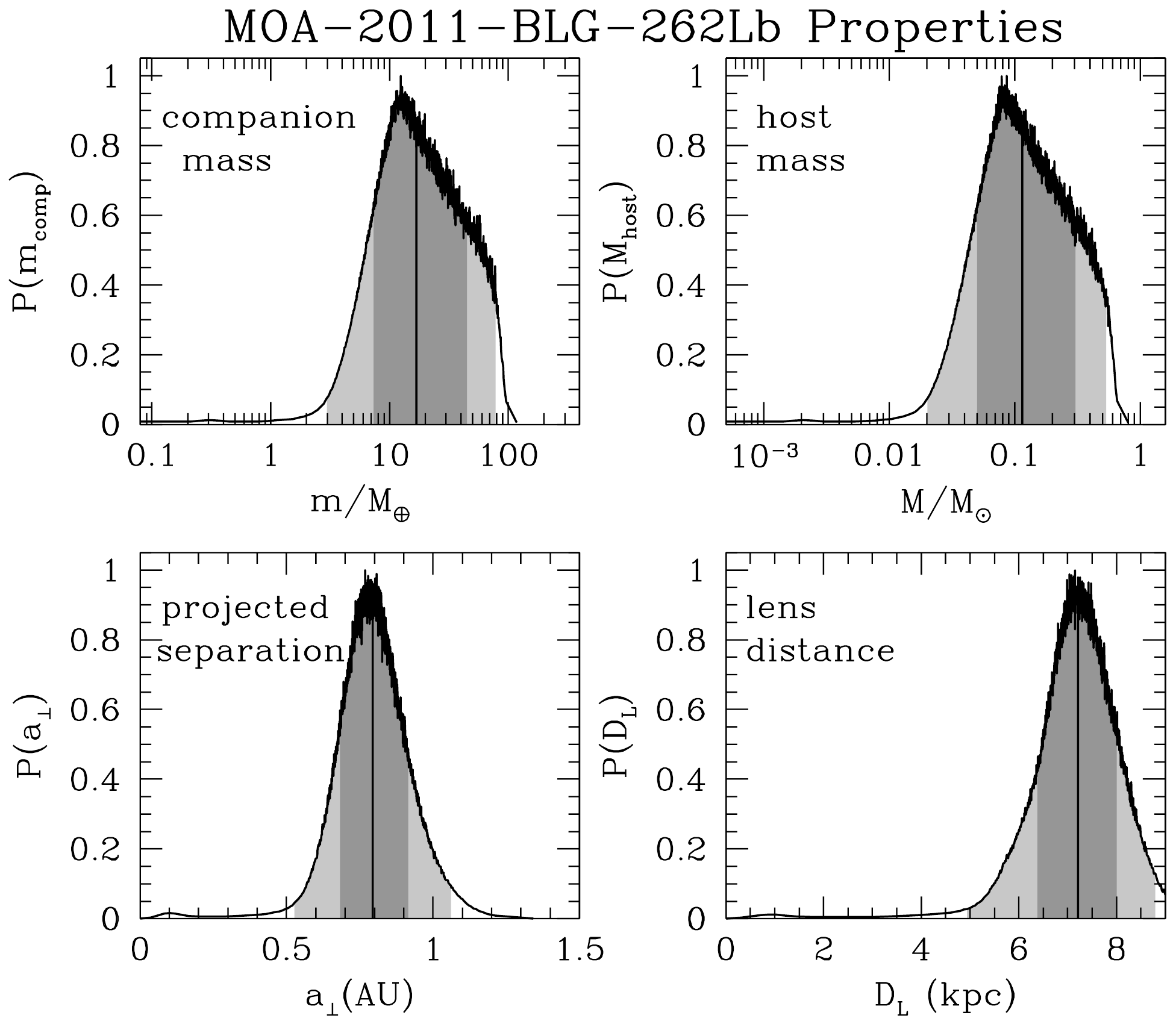}
\caption{Lens system properties for the wide and close versions of
the ``slow" solution only. This solution
clearly favors a low-mass star or high-mass brown dwarf host in the
Galactic bulge orbited by a planet of about a Neptune-mass. The lens
system would have a velocity that is significantly larger than is typical
of bulge stars.
\label{fig-lens_prop_slow}}
\end{figure}

The ``most likely" 
solution found by our Bayesian analysis for this fast solution
$M_L = 3.2 M_{\rm Jup}$ orbited by a moon of $m_m = 0.47\mearth$ at a 
3-d separation of $a = 0.13\,$AU, with the
lens system at a distance of $D_L = 0.56\,$kpc.
The slow solution has roughly equal probability (with our
prior assumptions) and implies a very different lens system.
With the slow solution, the host would most likely be a star with 
$M_L = 0.11{+0.21\atop -0.06}\msun$
orbited by a planet of mass $m_p = 17{+28\atop -10}\mearth$ at
a 3-d separation of $a = 0.95{+0.53\atop -0.19}\,$AU at a distance 
of $D_L = 7.2\pm 0.8\,$kpc, as shown in Figure~\ref{fig-lens_prop_slow}. 
These parameters are quite similar to the similar to the parameters 
for the secondary peak for the fast solution, shown in Figure~\ref{fig-lens_prop_fast},
except that the fast solution predicts a larger planet-star separation
(both the 3-d and projected separation).
Thus, the Bayesian analysis does not directly imply
a preferred solution. The probabilities are dependent
on our prior assumption about the probability of free-floating planets
to host a massive moon at a separation of $\sim 0.15\,$AU.
However, we do know that planets exist with similar parameters 
to the stellar mass host favored by the slow solution 
\citep{bennett08,moa192_naco,moa328}
and consistent with the fast solution, but we do not know of any
other planetary mass hosts with moons of about half an Earth-mass.
There are a handful of well sampled high magnification events 
with $t_E < 2\,$days, which implies that they probably have planetary mass primaries. 
For these events, MOA-ip-10 \citep{sumi11}, MOA-2009-BLG-450, and 
MOA-2010-BLG-418, we can largely exclude a companion with parameters
similar to MOA-2011-BLG-262Lb. This implies that planetary mass ratio
secondaries are not much more common around planetary mass hosts
than around stars and brown dwarfs. So, our Bayesian prior is not likely
to underestimate the probability of the planet+moon model by a significant
amount. On the other hand, such systems could be extremely rare, so 
out Bayesian prior could greatly overestimate the probability of such a system.
Or to put this another way, an apparently free-floating planet with a half
Earth-mass moon would be a new class of system that was not 
previously known to exist. Such a new discovery would require strong
evidence, so our favored model for this event is that it is a low-mass
star or brown dwarf orbited by a planet of about Neptune's mass.

Improvements in the observational capabilities in the near future
should allow the definitive detection of isolated planet+moon systems,
like the one favored by our Bayesian analysis for the best fit (fast)
model. Improved microlensing survey \citep{kmtnet} and follow-up
\citep{lcogt} capabilities, when coupled with the rapid event identification
demonstrated by the MOA group for this event, should allow the 
measurement of the terrestrial parallax effect \citep{gould09,yee09},
which would give a direct measurement of the lens system mass.
In the somewhat more distant future, a
future space-based microlensing survey \citep{bennett02}, such as the planned
WFIRST mission \citep{WFIRST_rep,WFIRST_AFTA} or a
microlensing program with the Euclid mission \citep{penny13}
will be even more sensitive to such events, and will enable the
detection of much lower mass lens primaries. Depending on the
details of the WFIRST orbit, this mission will also offer unique
ways to measure the microlensing parallax \citep{yee_WFIRST_par}
and therefore determine the lens masses. Thus, although
our conclusions regarding MOA-2011-BLG-262 are uncertain,
future microlensing surveys will be able to determine if there is
a large population of free-floating planets hosting Earth-mass moons.

\acknowledgments 
D.P.B.\ was supported by grants
NASA-NNX12AF54G, JPL-RSA \#1453175 and NSF AST-1211875.
This MOA project is supported by the grants JSPS18253002 and JSPS20340052.
T.S.\ acknowledges the financial support from the JSPS, JSPS23340044, JSPS24253004. 
This work was partially supported by a NASA Keck PI Data Award, administered by the NASA
Exoplanet Science Institute. Data presented herein were obtained at the W. M. Keck Observatory
from telescope time allocated to the National Aeronautics and Space Administration through the
agencyÕs scientific partnership with the California Institute of Technology and the University of
California. The Observatory was made possible by the generous financial support of the W. M.
Keck Foundation.
B.S.G.\ and A.G.\ were supported by NSF grant AST 110347.
B.S.G., A.G., R.P.G.\ were supported by NASA grant NNX12AB99G.
S.D.\ was partly supported through a Ralph E. and
Doris M. Hansmann Membership at the IAS and by NSF
grant AST-0807444.
Work by J.C.Y.\ was performed in part under contract with the California Institute 
of Technology (Caltech) funded by NASA through the Sagan Fellowship Program.
The OGLE project has received funding from the European Research
Council under the European Community's Seventh Framework Programme
(FP7/2007-2013) / ERC grant agreement no.\ 246678 to AU.
D.H.\ was supported by Czech Science Foundation grant GACR P209/10/1318.
D.M.B., M.D., K.H., C.S., R.A.S., M.H.\ and Y.T.\ are supported by NPRP grant 
NPRP-09-476-1-78 from the Qatar National Research Fund (a member of the Qatar Foundation).

\clearpage


\begin{figure}
\plotone{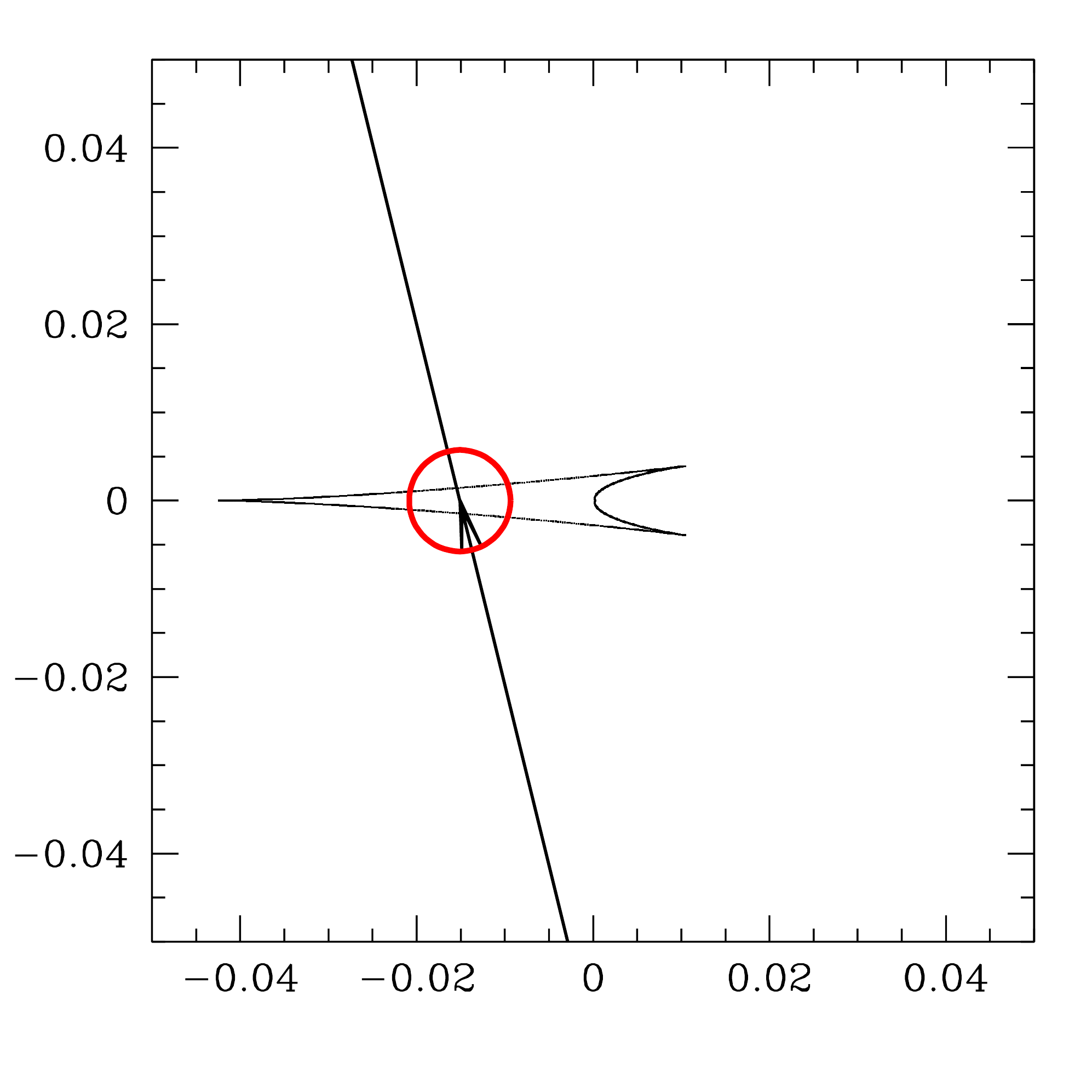}
\caption{ The caustic geometry for the best fit MOA-2011-BLG-262 ``slow" model.
The red circle indicates the source star size.
\label{fig-caustic_slow}}
\end{figure}

\clearpage

\begin{deluxetable}{cccccc}
\tablecaption{Model Parameters
                         \label{tab-mparams} }
\tablewidth{0pt}
\tablehead{
& & \multicolumn{2}{c} {fast} & \multicolumn{2}{c} {slow} \\
\colhead{parameter}  & \colhead{units} &
\colhead{$s<1$} & \colhead{$s>1$} & \colhead{$s<1$} & \colhead{$s>1$}
}  

\startdata

$t_E$ & days & 3.827 & 3.846 & 3.858 & 3.855 \\
$t_0$ & ${\rm HJD}-2455700$ & 39.1312 & 39.1311 & 39.1309 & 39.1310 \\
$u_{\rm min}$ & & 0.01465 & 0.01451 & 0.01470 & 0.01463 \\
$s$ & & 0.9578 & 1.0605 & 0.9263 & 1.0966 \\
$\theta$ & radians & 1.8096 & 1.8071 & 1.8109 & 1.8115 \\
$\epsilon$ & $10^{-4}$ & 4.661 & 4.674 & 4.391 & 4.434 \\
$t_\ast$ & days & 0.01316 & 0.01315 & 0.02217 & 0.02221 \\
$I_s$ & & 19.929 & 19.935 & 19.937 & 19.937 \\
$V_s$ & & 21.888 & 21.894 & 21.898 & 21.897 \\
fit $\chi^2$ &  & 5757.94 & 5758.58 & 5760.85  & 5763.82 \\

\enddata
\end{deluxetable}

\end{document}